\documentclass[journal]{IEEEtran}
\IEEEoverridecommandlockouts
\usepackage{lineno}
\usepackage{hyperref}
\usepackage{cite}
\usepackage{amsmath,amssymb,amsfonts,cases} 
\usepackage{amsmath}
\usepackage{amsthm}
\DeclareMathOperator*{\argmax}{argmax}

\usepackage{graphicx}
\usepackage{textcomp}
\usepackage{xcolor}
\usepackage{graphicx}
\usepackage{float}
\usepackage{subfigure}
\usepackage{amsmath,mleftright}
\usepackage{amsfonts,amssymb}
\usepackage{mathrsfs}
\usepackage{mathtools}
\usepackage{algorithm}
\usepackage{algorithmic}
\usepackage{bm}
\usepackage{multirow}
\usepackage{array}
\usepackage{amssymb}
\usepackage{amsmath}
\usepackage{cite}
\usepackage{url}
\usepackage{xcolor}
\usepackage{cite,graphicx,amsmath,amssymb}
\usepackage{subfigure}
\usepackage{fancyhdr}
\usepackage{mdwmath}
\usepackage{mdwtab}
\usepackage{caption}
\usepackage{amsthm}
\usepackage{setspace}
\usepackage{bm}
\usepackage{mathtools}
\usepackage{dsfont}
\usepackage{bbm}
\usepackage{framed}
\newtheorem{remark}{Remark}
\newtheorem{theorem}{Theorem}

\newtheorem{lemma}{Lemma}

\newtheorem{corollary}{Corollary}

\makeatletter
\newcommand{\biggg}{\bBigg@{3}}
\newcommand{\Biggg}{\bBigg@{3.5}}
\makeatother
\makeatletter
\renewcommand{\maketag@@@}[1]{\hbox{\m@th\normalsize\normalfont#1}}%
\makeatother
\def\BibTeX{{\rm B\kern-.05em{\sc i\kern-.025em b}\kern-.08em
    T\kern-.1667em\lower.7ex\hbox{E}\kern-.125emX}}
    \expandafter\def\expandafter\normalsize\expandafter{%
    \normalsize%
    \setlength\abovedisplayskip{4pt}%
    \setlength\belowdisplayskip{4pt}%
    \setlength\abovedisplayshortskip{2pt}%
    \setlength\belowdisplayshortskip{2pt}%
}
\begin{document}
\title{Rate Region of ISAC for Pinching-Antenna Systems}
\author{Chongjun~Ouyang, Zhaolin~Wang, Yuanwei~Liu, and Zhiguo~Ding\vspace{-10pt}
\thanks{C. Ouyang and Z. Wang are with the School of Electronic Engineering and Computer Science, Queen Mary University of London, London, E1 4NS, U.K. (e-mail: \{c.ouyang, zhaolin.wang\}@qmul.ac.uk).}
\thanks{Y. Liu is with the Department of Electrical and Electronic Engineering, The University of Hong Kong, Hong Kong (email: yuanwei@hku.hk).}
\thanks{Z. Ding is with the School of Electrical and Electronic Engineering, The University of Manchester, Manchester, M13 9PL, U.K., and also with the Department of Electrical Engineering and Computer Science, Khalifa University, Abu Dhabi, UAE (e-mail: zhiguo.ding@manchester.ac.uk).}}
\maketitle
\begin{abstract}
The Pinching-Antenna SyStem (PASS) reconstructs wireless channels through \emph{pinching beamforming}, wherein the activated positions of pinching antennas along dielectric waveguides are optimized to shape the radiation pattern. The aim of this article is to analyze the performance limits of employing PASS in integrated sensing and communications (ISAC). Specifically, a PASS-assisted ISAC system is considered, where a pinched waveguide is utilized to simultaneously communicate with a user and sense a target. Closed-form expressions for the achievable communication rate (CR) and sensing rate (SR) are derived to characterize the information-theoretic limits of this dual-functional operation. \romannumeral1) For the single-pinch case, closed-form solutions for the optimal pinching antenna location are derived under \emph{sensing-centric (S-C)}, \emph{communications-centric (C-C)}, and \emph{Pareto-optimal} designs. On this basis, the CR-SR trade-off is characterized by deriving the full CR-SR rate region, which is shown to encompass that of conventional fixed-antenna systems. \romannumeral2) For the multiple-pinch case, an antenna location refinement method is applied to obtain the optimal C-C and S-C pinching beamformers. As a further advance, inner and outer bounds on the achievable CR-SR region are derived using an element-wise alternating optimization technique and by invoking Cauchy-Schwarz and Karamata's inequalities, respectively. Numerical results demonstrate that: \romannumeral1) the derived bounds closely approximate the true CR-SR region; and \romannumeral2) PASS can achieve a significantly larger rate region than conventional-antenna systems.
\end{abstract}
\section{Introduction}
Since Paulraj and Kailath patented the multiple-input multiple-output (MIMO) concept \cite{paulraj1994increasing} and Foschini established its technical foundations in fading channels \cite{foschini1996layered} during the 1990s, multiple-antenna technology has reshaped and revolutionized modern wireless communication systems \cite{heath2018foundations}. Over the past decades, this technology has evolved along two notable directions \cite{bjornson2019massive,liu2024near}. The first involves increasing the array aperture size and the density of element deployment, which has led to the development of massive MIMO \cite{larsson2014massive}, gigantic MIMO \cite{li2023gigantic}, and holographic MIMO \cite{hu2018beyond,pizzo2020spatially}. The second focuses on flexible array architectures capable of reconfiguring the wireless channel. This line of research leads to flexible-antenna systems, including reconfigurable intelligent surfaces (RIS) \cite{liu2021reconfigurable} as well as more recent innovations including fluid antennas \cite{wong2020fluid} and movable antennas \cite{zhu2024movable}.

Despite their advancements, these evolved MIMO architectures have limited effectiveness in addressing fundamental issues, namely free-space path loss and line-of-sight (LoS) blockage, which are two primary causes of signal impairments in wireless propagation. Enlarging the aperture and increasing the array density can enhance array gains and spatial degrees of freedom, but these approaches fall short in mitigating LoS blockage and free-space path loss, especially for users at the cell edge. The RIS offers a potential solution by creating virtual LoS links between transceivers. However, they suffer from significant double-fading effects due to electromagnetic (EM) wave reflection \cite{ozdogan2019intelligent}. Fluid and movable antennas aim to improve channel conditions by optimizing antenna positions, but their movements are typically limited to apertures on the order of a few wavelengths. This constraint reduces their effectiveness in overcoming large-scale path loss and LoS obstruction. Moreover, once deployed, these systems generally do not allow for flexible changes in antenna configuration, such as adding or removing antenna elements, without incurring significant cost or complexity.

\begin{figure}[!t]
\centering
\includegraphics[width=0.4\textwidth]{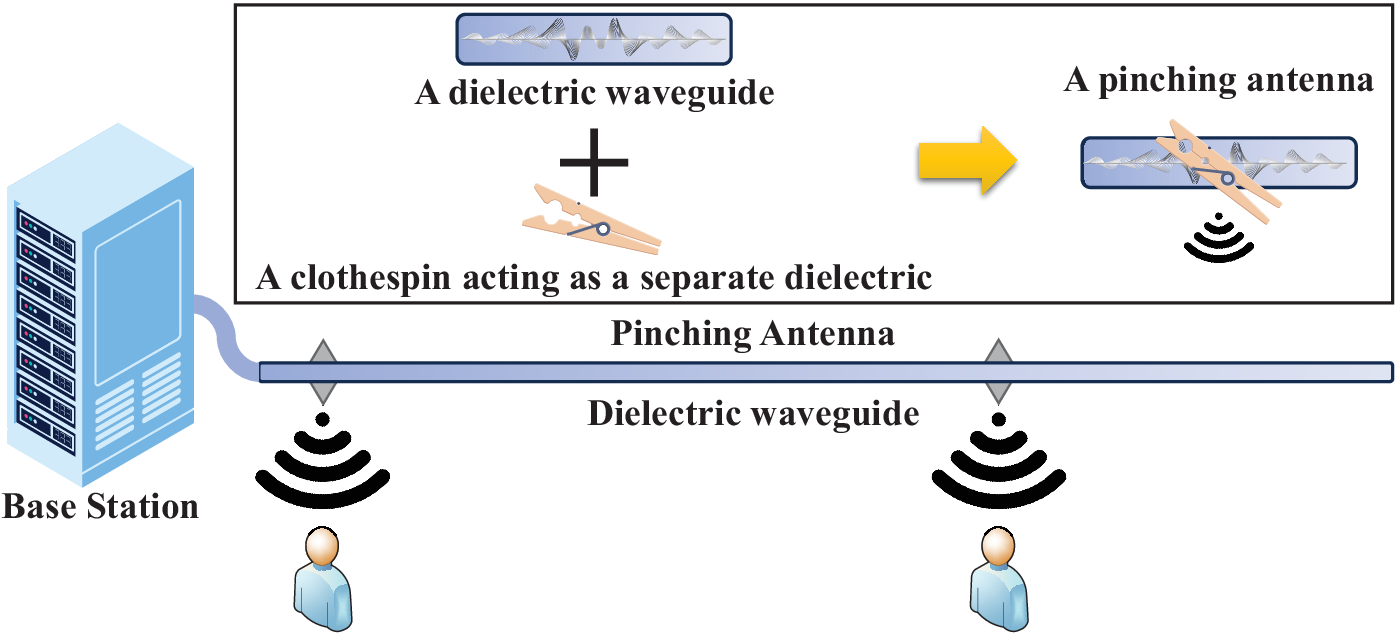}
\caption{Illustration of a PASS.}
\label{Figure_PASS}
\vspace{-10pt}
\end{figure}

To overcome the aforementioned limitations, NTT DOCOMO has introduced the \emph{pinching antenna} \cite{pinching_antenna1,suzuki2022pinching}, a novel form of flexible-antenna technology. This approach employs a dielectric waveguide \cite{pozar2021microwave} as the transmission medium, where EM waves are emitted by attaching small dielectric particles at designated locations along the waveguide \cite{suzuki2022pinching}. These dielectric elements are typically mounted on the tips of plastic clips, forming what is referred to as pinching antennas \cite{yang2025pinching,liu2025pinching}. Each pinching antenna can be independently activated or deactivated at any point along the waveguide. This capability enables dynamic reconfiguration of the antenna array \cite{pinching_antenna1,suzuki2022pinching}, much like attaching or removing clothespins on a clothesline, as illustrated in {\figurename} {\ref{Figure_PASS}}. This mechanism facilitates a highly flexible and scalable deployment strategy, which we refer to as \emph{pinching beamforming} \cite{liu2025pinching}. 

Unlike conventional flexible-antenna systems, the waveguide in the \emph{Pinching-Antenna SyStem (PASS)} can be extended to arbitrary lengths, which allows antennas to be positioned in close proximity to users to establish robust LoS links. A PASS is also cost-effective and easy to install, as it requires only the addition or removal of dielectric materials. In essence, PASS can be regarded as a practical realization of existing flexible-antenna systems \cite{new2024tutorial,zhu2025tutorial}, but with enhanced flexibility and scalability. In recognition of NTT DOCOMO's original contribution \cite{pinching_antenna1,suzuki2022pinching}, we refer to this technology as \emph{PASS} throughout this article.
\subsection{Prior Works}
Owing to its distinctive characteristics, PASS-assisted communications has attracted growing research interest. A pioneering study by Ding \emph{et al.} examined the average transmission rate performance achieved through pinching antennas serving mobile users and provided a theoretical comparison with conventional fixed-position antenna systems \cite{ding2024flexible}. Subsequent work analyzed the outage probability of PASS while incorporating hardware impairments such as in-waveguide propagation loss \cite{tyrovolas2025performance}. The array gain of PASS was also investigated to determine the optimal number of pinching antennas for deployment \cite{ouyang2025array}. A fundamental power radiation model was later developed based on EM coupling theory \cite{wang2025modeling}, and the impact of LoS blockage on PASS performance was further explored \cite{ding2025blockage}. Collectively, these studies confirmed that PASS effectively mitigates large-scale path loss and LoS blockage, and outperforms traditional fixed-antenna and existing fluid/movable-antenna systems. Building on this theoretical groundwork, several pinching beamforming algorithms have been proposed to optimize the placement of active pinching antennas along the waveguide, so as to enhance the system's overall communication performance \cite{xu2024rate,wang2024antenna,tegos2024minimum,wang2025modeling,bereyhi2025downlink,bereyhi2025mimo}.

In addition to the recent advances in PASS-based communications, efforts have begun exploring the application of PASS in wireless sensing and integrated sensing and communications (ISAC), one of the six key usage scenarios for the sixth-generation (6G) networks as defined by ITU-R IMT-2030 \cite{itu2023framework}. The Cram\'{e}r-Rao lower bounds for PASS-assisted user-centric localization \cite{ding2025pinching} and array-centric localization \cite{bozanis2025CRLB} have been analyzed. Beyond localization, PASS has also been investigated for other sensing tasks such as target tracking \cite{qin2025joint}, target detection \cite{zhang2025integrated}, and radar cross-section (RCS) sensing \cite{khalili2025pinching}, where these sensing functions have been jointly optimized with communication performance. These initial studies laid the groundwork for applying PASS in sensing and ISAC, highlighting its advantages in terms of low cost and high reconfigurability. 
\subsection{Motivation and Contributions}
Despite promising early results, research on the application of PASS in ISAC is still in its infancy. Existing works \cite{qin2025joint,zhang2025integrated,khalili2025pinching} have primarily focused on the optimization perspective, i.e., formulating dual-functional sensing and communication (DFSAC) pinching beamforming problems and proposing algorithmic solutions. However, these efforts have not established the fundamental performance limits of PASS-enabled ISAC systems. In particular, they do not fully characterize the crucial tradeoff between communication and sensing performance.

To address this knowledge gap and advance the theoretical understanding of PASS in ISAC applications, this paper analyzes the fundamental limits of an ISAC system equipped with pinching antennas. We study how pinching beamforming affects the tradeoff between communications and sensing by characterizing the achievable \emph{information-theoretic rate region} \cite{ouyang2023integrated}. The main contributions are summarized as follows.
\begin{itemize}
  \item First, we develop a PASS-enabled ISAC framework in which a base station (BS) employs a single pinched dielectric waveguide to communicate with a user and sense a target simultaneously. Within this framework, we derive closed-form expressions for the achievable communication rate (CR) and sensing rate (SR) in order to quantify the information-theoretic limits of both tasks \cite{ouyang2023integrated}. We demonstrate that these metrics depend explicitly on the pinching beamformer, i.e., the locations of the pinching antennas, which enables the design of DFSAC pinching beamforming strategies.
  \item In order to obtain insightful understandings, we focus on the single-pinch case, where a single pinching antenna is applied to the waveguide. In this scenario, we consider three pinching beamforming design objectives to characterize the tradeoff between CR and SR: \romannumeral1) a sensing-centric (S-C) design to maximize SR, \romannumeral2) a communications-centric (C-C) design to maximize CR, and \romannumeral3) a Pareto-optimal design to characterize the full CR-SR rate region. For each objective, we provide closed-form solutions for the optimal activated location of the pinching antenna and prove that the rate region of conventional-antenna system is nested within the region of PASS.
  \item Next, we extend the analysis to a more general multiple-pinch case, where multiple pinching antennas are deployed along the DFSAC waveguide. We employ an antenna location refinement method to optimize the activated locations of the pinching antennas for both C-C and S-C designs. Because obtaining the Pareto-optimal solution is analytically intractable, an element-wise alternating optimization method is combined with time sharing in order to compute an \emph{achievable inner bound} of the CR-SR rate region. Additionally, we derive a closed-form \emph{outer bound} by leveraging the Cauchy-Schwarz and Karamata's inequalities.
  \item Finally, we present numerical simulations to verify the theoretical analysis and examine the tightness of the proposed bounds. The results demonstrate that \romannumeral1) the inner and outer bounds closely approximate the true CR-SR region in the multiple-pinch case, \romannumeral2) PASS significantly outperforms conventional-antenna systems in both CR and SR, achieving a larger rate region, and \romannumeral3) increasing the number of pinching antennas can enlarge the achievable rate region. These findings highlight the scalability and effectiveness of PASS in ISAC applications.
\end{itemize}

The remainder of this paper is organized as follows. Section {\ref{Section: System Model}} introduces the system model for the PASS-enabled ISAC system. Sections \ref{Section: Single-Pinch Case} and \ref{Section: Multiple-Pinch Case} analyze the rate regions achieved in the single-pinch and multiple-pinch cases, respectively. Section \ref{Section_Numerical_Results} presents numerical results along with detailed discussions. Finally, Section \ref{Section_Conclusion} concludes the paper.
\subsubsection*{Notations}
Throughout this paper, scalars, vectors, and matrices are denoted by non-bold, bold lower-case, and bold upper-case letters, respectively. For a matrix $\mathbf{A}$, ${\mathbf{A}}^{\mathsf{T}}$ and ${\mathbf{A}}^{\mathsf{H}}$ denote the transpose and conjugate transpose of $\mathbf{A}$, respectively. For a square matrix $\mathbf{B}$, ${\mathbf{B}}^{-1}$ and $\det(\mathbf{B})$ denote the inverse and determinant of $\mathbf{B}$, respectively. The notations $\lvert a\rvert$ and $\lVert \mathbf{a} \rVert$ represent the magnitude of scalar $a$ and the norm of vector $\mathbf{a}$, respectively. The identity matrix of size $N\times N$ is denoted by $\mathbf{I}_N$, and the all-zero matrix with appropriate dimension is represented by $\mathbf{0}$. The sets $\mathbbmss{C}$ and $\mathbbmss{R}$ denote the complex and real spaces, respectively. The floor operator is denoted by $\lfloor\cdot\rfloor$, and $\mathbbmss{E}\{\cdot\}$ denotes statistical expectation. The notation ${\mathcal{CN}}({\bm\mu},\mathbf{X})$ refers to the circularly symmetric complex Gaussian distribution with mean $\bm\mu$ and covariance matrix $\mathbf{X}$. The convex hull of a set is denoted by $\rm{Conv}(\cdot)$, and the union operation is denoted by $\bigcup$.
\section{System Model}\label{Section: System Model}
\begin{figure}[!t]
\centering
\includegraphics[width=0.45\textwidth]{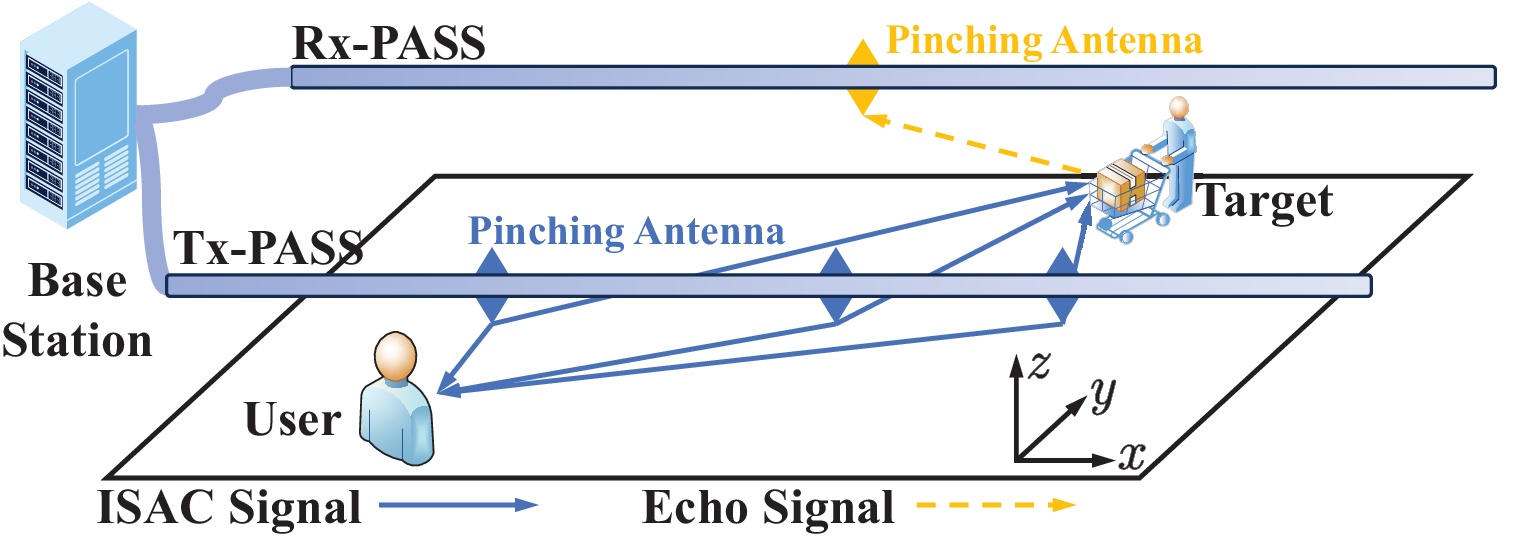}
\caption{Illustration of an ISAC system with pinching antennas.}
\label{Figure1}
\vspace{-10pt}
\end{figure}

In an ISAC system as illustrated in {\figurename} {\ref{Figure1}}, a BS simultaneously communicates with a single-antenna user and senses a single target. The BS is equipped with two dielectric waveguides for transmission and reception, respectively. On the \emph{transmit waveguide (Tx-PASS)}, $N$ pinching antennas are activated to jointly facilitate both communication and sensing tasks, as shown in {\figurename} {\ref{Figure1}}. Meanwhile, the \emph{receive waveguide (Rx-PASS)} is used to collect the sensing echo reflected by the target and forward it back to the BS for further processing. Since the system focuses on sensing a single target, a single activated pinching antenna is sufficient, which is also useful to obtain insightful analytical results. 

Let ${\mathbf{x}}\triangleq[x_1,\ldots,x_L]\in{\mathbbmss{C}}^{1\times L}$ denote the DFSAC signal vector, where $L$ is the length of the communication frame or the number of sensing pulses. From a communication standpoint, $x_{\ell}\in{\mathbbmss{C}}$ denotes the $\ell$th data symbol for $\ell\in\{1,\ldots,L\}$. For sensing, $x_\ell$ represents the snapshot of the probing signal at the $\ell$th time slot. In the considered ISAC system, we have ${\mathbf{x}}=\sqrt{P}{\mathbf{s}}$, where $P$ is the power budget, and ${\mathbf{s}}\in{\mathbbmss{C}}^{1\times L}$ denotes the unit-power data stream intended for the communication user with $L^{-1}\lVert{\mathbf{s}}\rVert^2=1$.

Recall that most applications of ISAC and PASS are for the scenario with high-frequency bands \cite{suzuki2022pinching}, where LoS propagation tends to dominate \cite{ouyang2024primer}. Therefore, we adopt a free-space LoS channel model to theoretically examine the performance limits of PASS-enabled ISAC systems. The influence of multipath fading is beyond the scope of this work and will be considered in future investigations.
\subsection{Communication Model with PASS}
The transmit waveguide is aligned parallel to the $x$-axis at a height $d$. The spatial channel coefficient between the $n$th pinching antenna and the user is expressed as follows \cite{liu2023near}:
\begin{align}
h({\mathbf{u}}_{\rm{c}},{\bm\psi}_n)\triangleq\frac{\eta^{\frac{1}{2}}{\rm{e}}^{-{\rm{j}}k_0\lVert{\mathbf{u}}_{\rm{c}}-{\bm\psi}_n\rVert}}{\lVert{\mathbf{u}}_{\rm{c}}-{\bm\psi}_n\rVert},
\end{align}
where $\eta\triangleq\frac{c^2}{16\pi^2f_{\rm{c}}^2}$, $c$ denotes the speed of light, $f_{\rm{c}}$ is the carrier frequency, $k_0=\frac{2\pi}{\lambda}$ is the wavenumber, and $\lambda$ is the free-space wavelength. Furthermore, ${\mathbf{u}}_{\rm{c}}\triangleq[x_{\rm{c}},y_{\rm{c}},0]^{\mathsf{T}}$ denotes the location of the user, and ${\bm\psi}_n\triangleq[t_n,y_{\rm{t}},d]^{\mathsf{T}}$ denotes the location of the $n$th pinching antenna for $n\in{\mathcal{N}}\triangleq\{1,\ldots,N\}$. Without loss of generality, we consider 
\begin{align}
t_{\max}\geq t_{n}>t_{n'},~\forall n>n', 
\end{align}
where $t_{\max}$ denotes the maximum deployment range of the pinching antennas. 

The received signal at the user is given by
\begin{subequations}\label{Signal_Model_Scalar}
\begin{align}
{\mathbf{y}}_{\rm{c}}&=\sum_{n=1}^{N}\sqrt{{P}/{N}}h({\mathbf{u}}_{\rm{c}},{\bm\psi}_n){\rm{e}}^{-{\rm{j}}{\phi_n}}{\mathbf{s}}+{\mathbf{z}}_{\rm{c}}\\
&=\sqrt{{P}/{N}}{\mathbf{h}}^{\mathsf{T}}({\mathbf{u}}_{\rm{c}},{\mathbf{t}}){\bm\phi}_{\rm{t}}{\mathbf{s}}+{\mathbf{z}}_{\rm{c}}\in{\mathbbmss{C}}^{1\times L},
\end{align}
\end{subequations}
where 
\begin{subequations}
\begin{align}
{\bm\phi}_{\rm{t}}&\triangleq[{\rm{e}}^{-{\rm{j}}\phi_{1}},\ldots,{\rm{e}}^{-{\rm{j}}{\phi_{N}}}]^{\mathsf{T}}\in{\mathbbmss{C}}^{N\times1},\\
{\mathbf{h}}({\mathbf{u}}_{\rm{c}},{\mathbf{t}})&\triangleq[h({\mathbf{u}}_{\rm{c}},{\bm\psi}_1),\ldots,h({\mathbf{u}}_{\rm{c}},{\bm\psi}_N)]^{\mathsf{T}}\in{\mathbbmss{C}}^{N\times1},\\
{\mathbf{t}}&\triangleq[t_1,\ldots,t_N]^{\mathsf{T}}\in{\mathbbmss{R}}^{N\times1},
\end{align}
\end{subequations}
and ${\mathbf{z}}_{\rm{c}}\in{\mathbbmss{C}}^{1\times L}$ is additive Gaussian noise with ${\mathbf{z}}_{\rm{c}}^{\mathsf{H}}\sim{\mathcal{CN}}({\mathbf{0}},\sigma_{\rm{c}}^2{\mathbf{I}}_L)$ and $\sigma_{\rm{c}}^2$ being the noise power. The overall transmit power $P$ is assumed to be equally distributed among the $N$ active pinching antennas \cite{ding2024flexible,wang2025modeling}, which results in the power allocation coefficient $\frac{P}{N}$ as shown in \eqref{Signal_Model_Scalar}. The term 
\begin{align}
\phi_n\triangleq\frac{2\pi\lVert{\bm\psi}_n-{\bm\psi}_0\rVert}{\lambda_{\rm{g}}}=\frac{2\pi\lvert t_n-t_0\rvert}{\lambda_{\rm{g}}} 
\end{align}
denotes the in-waveguide phase shift with respect to the $n$th pinching antenna for $n\in\mathcal{N}$, where ${\bm\psi}_0\triangleq[t_0,y_{\rm{t}},d]^{\mathsf{T}}$ represents the location of the waveguide's feed point with $t_0\leq t_{1}$, $\lambda_{\rm{g}}=\frac{\lambda}{n_{\rm{eff}}}$ is the guided wavelength, and $n_{\rm{eff}}$ is the effective refractive index of the dielectric waveguide \cite{pozar2021microwave}. 

In-waveguide propagation loss is \emph{neglected} in this model, as its impact on the overall system performance is insignificant due to the dominance of free-space path loss, as demonstrated in prior work \cite{ding2024flexible,ouyang2025array,wang2024antenna}. As a result, the performance results obtained in this study can be interpreted as upper bounds for ISAC systems employing pinching antennas. However, the impact of in-waveguide propagation loss will be further examined in the simulation section. 

According to \eqref{Signal_Model_Scalar}, the user's signal-to-noise ratio (SNR) for decoding ${\mathbf{s}}$ can be written as follows:
\begin{subequations}
\begin{align}
\gamma_{\rm{c}}&\triangleq\frac{P}{\sigma_{\rm{c}}^2}\frac{\lvert{\mathbf{h}}^{\mathsf{T}}({\mathbf{u}}_{\rm{c}},{\mathbf{t}}){\bm\phi}_{\rm{t}}\rvert^2}{N}\\
&=\overline{\gamma}_{\rm{c}}\left\lvert\sum_{n=1}^{N}\frac{{{\rm{e}}^{-{\rm{j}}k_0\sqrt{d_{\rm{c}}^2+(t_n-x_{\rm{c}})^2}-{\rm{j}}{k_0(t_n-t_{0})}n_{\rm{eff}}}}}
{\sqrt{d_{\rm{c}}^2+(t_n-x_{\rm{c}})^2}}\right\rvert^2,\label{Communication_SNR_Basic_Expression}
\end{align}
\end{subequations}
with $\overline{\gamma}_{\rm{c}}\triangleq\frac{P\eta}{N\sigma_{\rm{c}}^2}$ and $d_{\rm{c}}^2\triangleq(y_{\rm{c}}-y_{\rm{t}})^2+d^2$. Based on this, the CR is given by
\begin{align} 
{\mathcal{R}}_{\rm{c}}=\log_2(1+\gamma_{\rm{c}})
=\log_2\left(1+\frac{P}{\sigma_{\rm{c}}^2}\frac{\lvert{\mathbf{h}}^{\mathsf{T}}({\mathbf{u}}_{\rm{c}},{\mathbf{t}}){\bm\phi}_{\rm{t}}\rvert^2}{N}\right).
\end{align}
\subsubsection*{Conventional Fixed-Antenna System}
For comparison, we also present the CR achieved by a conventional fixed-antenna system, where a single transmit antenna\footnote{We assume that the conventional-antenna system employs only a single transmit antenna, not only due to the high cost of deploying additional antennas, but also because such systems inherently lack such flexibility afforded by the PASS architecture.} is located at $[t_{\rm{f}},y_{\rm{t}},d]^{\mathsf{T}}$. The corresponding CR is given by
\begin{align}\label{CASS_CR}
{\mathcal{R}}_{\rm{c}}^{\rm{f}}\triangleq\log_2\left(1+\frac{\overline{\gamma}_{\rm{c}}^{\rm{f}}}{d_{\rm{c}}^2+(t_{\rm{f}}-x_{\rm{c}})^2}\right),
\end{align}
where ${\overline{\gamma}_{\rm{c}}^{\rm{f}}}\triangleq\frac{P\eta}{\sigma_{\rm{c}}^2}$. By comparing \eqref{CASS_CR} with the CR expression under PASS, we observe that the CR achieved by PASS depends on the activated locations of the pinching antennas, or equivalently, the \emph{pinching beamformer} $\mathbf{t}=[t_1,\ldots,t_N]^{\mathsf{T}}$. This additional spatial flexibility enables PASS to potentially establish a more stable and stronger LoS link to the user compared to the conventional fixed-antenna system, which lacks the ability to dynamically reconfigure or reconstruct its wireless channel.
\subsection{Sensing Model with PASS}
The receive waveguide is positioned parallel to the $x$-axis at a height $d$, where the location of the receive pinching antenna is denoted as ${\bm\varphi}\triangleq[r,y_{\rm{r}},d]^{\mathsf{T}}$. When the system transmits signal ${\mathbf{x}}$ to sense the target located at ${\mathbf{u}}_{\rm{s}}\triangleq[x_{\rm{s}},y_{\rm{s}},0]^{\mathsf{T}}$, the waveguide receives the following reflected echo signal:
\begin{align}
{\mathbf{y}}_{\rm{s}}=\sqrt{{1}/{N}}{\mathbf{g}}^{\mathsf{T}}{\bm\phi}_{\rm{t}}{\mathbf{x}}+{\mathbf{n}}_{\rm{s}}\in{\mathbbmss{C}}^{1\times L},
\end{align}
where ${\mathbf{n}}_{\rm{s}}\in{\mathbbmss{C}}^{1\times L}$ is the noise vector with zero-mean entries and variance $\sigma_{\rm{s}}^2$, and ${\mathbf{g}}\in{\mathbbmss{C}}^{1\times N}$ represents the target response vector. This vector can be modeled as follows \cite{tang2018spectrally,liu2024road}:
\begin{align}
{\mathbf{g}}={\beta}_{\rm{RCS}}{{g}}_{\rm{r}}{\mathbf{h}}^{\mathsf{T}}({\mathbf{u}}_{\rm{s}},{\mathbf{t}}),
\end{align}
where ${\beta}_{\rm{RCS}}\in{\mathbbmss{C}}$ denotes the RCS of the sensing target. The channel response from the target to the receive pinching antenna is given by 
\begin{align}\label{Target_Response_Receive}
{{g}}_{\rm{r}}\triangleq h({\mathbf{u}}_{\rm{s}},{\bm\varphi})
=\frac{\eta^{\frac{1}{2}}{\rm{e}}^{-{\rm{j}}k_0\sqrt{d_{\rm{r}}^2+(x_{\rm{s}}-r)^2}}}{\sqrt{d_{\rm{r}}^2+(x_{\rm{s}}-r)^2}} 
\end{align}
with $d_{\rm{r}}^2\triangleq{d^2+(y_{\rm{s}}-y_{\rm{r}})^2}$, and ${\mathbf{h}}({\mathbf{u}}_{\rm{s}},{\mathbf{t}})\in{\mathbbmss{C}}^{N\times 1}$ denotes the transmit beam-steering vector. Without loss of generality, we assume $x_{\rm{c}}\leq x_{\rm{s}}$.

The signal ${\mathbf{y}}_{\rm{s}}$ is then passed through the waveguide, and the corresponding observation at the BS is given by
\begin{subequations}\label{Sensing Model}
\begin{align}
{\phi}_{\rm{r}}{\mathbf{y}}_{\rm{s}}&=
\sqrt{{1}/{N}}{\beta}_{\rm{RCS}}{\phi}_{\rm{r}}{{g}}_{\rm{r}}{\mathbf{h}}^{\mathsf{T}}({\mathbf{u}}_{\rm{s}},{\mathbf{t}}){\bm\phi}_{\rm{t}}{\mathbf{x}}
+{\phi}_{\rm{r}}{\mathbf{n}}_{\rm{s}}\\
&=\sqrt{{P}/{N}}{\beta}_{\rm{RCS}}{\phi}_{\rm{r}}{{g}}_{\rm{r}}{\mathbf{h}}^{\mathsf{T}}({\mathbf{u}}_{\rm{s}},{\mathbf{t}}){\bm\phi}_{\rm{t}}{\mathbf{s}}
+{\phi}_{\rm{r}}{\mathbf{n}}_{\rm{s}}.
\end{align}
\end{subequations}
where 
\begin{align}
{\phi}_{\rm{r}}\triangleq{\rm{e}}^{-{\rm{j}}\frac{2\pi\lVert{\bm\varphi}-{\bm\varphi}_0\rVert}{\lambda_{\rm{g}}}}
={\rm{e}}^{-{\rm{j}}\frac{2\pi(r-r_0)}{\lambda_{\rm{g}}}},
\end{align}
with ${\bm\varphi}_0\triangleq[r_0,y_{\rm{r}},d]^{\mathsf{T}}$ representing the location of the feed point of the waveguide with $r_0\leq r$. As a result, ${\phi}_{\rm{r}}{\mathbf{n}}_{\rm{s}}^{\mathsf{T}}\sim{\mathcal{CN}}({\mathbf{0}},\sigma_{\rm{s}}^2{\mathbf{I}}_L)$. The BS uses the observation ${\phi}_{\rm{r}}{\mathbf{y}}_{\rm{s}}$ to perform sensing. We assume that the location of the target is perfectly tracked, and focus on estimating the RCS ${\beta}_{\rm{RCS}}$ \cite{liu2024road}. The estimated RCS can subsequently be used to reconstruct the EM properties of the target, such as its relative permittivity and conductivity, which enables high-resolution environmental imaging \cite{jiang2024electromagnetic,ouyang2023integrated}. 

Following the \emph{Swerling-\uppercase\expandafter{\romannumeral1}} model \cite{richards2005fundamentals}, we assume the RCS is relatively constant from pulse-to-pulse with an \emph{a prior} Rayleigh distributed amplitude, resulting in ${\beta}_{\rm{RCS}}\sim{\mathcal{CN}}(0,\alpha_{\rm{s}})$, where $\alpha_{\rm{s}}$ denotes the average reflection strength of the target \cite{richards2005fundamentals}. The information-theoretic limits on this sensing task is characterized by the sensing mutual information (MI), which is defined as the MI between ${\phi}_{\rm{r}}{\mathbf{y}}_{\rm{s}}$ and ${\beta}_{\rm{RCS}}$ (or $\mathbf{g}$), conditioned on $\mathbf{x}$ \cite{ouyang2023integrated,tang2018spectrally,liu2024road}. Based on this, we adopt the SR as the metric, which quantifies the sensing MI per unit time. Assuming each DFSAC symbol occupies one unit of time, the SR is defined as follows \cite{ouyang2023integrated}:
\begin{align}
{\mathcal{R}}_{\rm{s}}=L^{-1}I({\phi}_{\rm{r}}{\mathbf{y}}_{\rm{s}};{\mathbf{g}}|{\mathbf{x}}),
\end{align}
where $I(X;Y|Z)$ denotes the MI between $X$ and $Y$ conditioned on $Z$. In particular, ${\mathcal{R}}_{\rm{s}}$ can be calculated in the following form.
\vspace{-5pt}
\begin{lemma}\label{Lemma_SR}
Given $\{{\phi}_{\rm{r}},{{g}}_{\rm{r}},{\mathbf{h}}({\mathbf{u}}_{\rm{s}},{\mathbf{t}}),{\bm\phi}_{\rm{t}}\}$, the SR achieved by the considered PASS can be expressed as follows:
\begin{align}\label{Lemma_SR_Exp}
{\mathcal{R}}_{\rm{s}}=L^{-1}\log_2(1+\gamma_{\rm{s}}),
\end{align}
where $\gamma_{\rm{s}}\triangleq\frac{PL\alpha_{\rm{s}}}
{N\sigma_{\rm{s}}^2}\lvert{{g}}_{\rm{r}}\rvert^2\rvert{\mathbf{h}}^{\mathsf{T}}({\mathbf{u}}_{\rm{s}},{\mathbf{t}})
{\bm\phi}_{\rm{t}}\rvert^2$ is defined as the effective sensing SNR.
\end{lemma}
\vspace{-5pt}
\begin{IEEEproof}
Refer to Appendix \ref{Proof_Lemma_SR} for more details.
\end{IEEEproof}
Furthermore, the following lemma establishes the relationship between the SR, or equivalently the sensing MI, and the estimation error in recovering the RCS ${\beta}_{\rm{RCS}}$.
\vspace{-5pt}
\begin{lemma}\label{Lemma_MSE_SR}
Under the considered sensing model, maximizing the achievable SR is equivalent to minimizing the mean-squared error (MSE) in estimating the RCS.
\end{lemma}
\vspace{-5pt}
\begin{IEEEproof}
Refer to Appendix \ref{Proof_MSE_SR} for more details.
\end{IEEEproof}
Upon referring to \eqref{Target_Response_Receive} and \eqref{Lemma_SR_Exp}, we find that in order to maximize the SR, we need to set the location of the receive pinching antenna aligned with the projection of the target's location on the waveguide, i.e., $r=x_{\rm{s}}$. In doing so, we have
\begin{align}
\gamma_{\rm{s}}=\frac{P\rvert{\mathbf{h}}^{\mathsf{T}}({\mathbf{u}}_{\rm{s}},{\mathbf{t}})
{\bm\phi}_{\rm{t}}\rvert^2L\alpha_{\rm{s}}}
{N\sigma_{\rm{s}}^2}\frac{\eta}{{d^2+(y_{\rm{s}}-y_{\rm{r}})^2}}.
\end{align}

Letting $\overline{\gamma}_{\rm{s}}\triangleq\frac{PL\eta^2\alpha_{\rm{s}}}
{N\sigma_{\rm{s}}^2d_{\rm{r}}^2}$, we can rewrite $\gamma_{\rm{s}}$ as follows:
\begin{align}\label{Sensing_SNR_Basic_Expression}
\gamma_{\rm{s}}=
\overline{\gamma}_{\rm{s}}\left\lvert\sum_{n=1}^{N}\frac{{{\rm{e}}^{-{\rm{j}}k_0\sqrt{d_{\rm{s}}^2+(t_n-x_{\rm{s}})^2}-{\rm{j}}{k_0(t_n-t_{0})}n_{\rm{eff}}}}}
{\sqrt{d_{\rm{s}}^2+(t_n-x_{\rm{s}})^2}}\right\rvert^2,
\end{align}
with $d_{\rm{s}}^2\triangleq(y_{\rm{s}}-y_{\rm{t}})^2+d^2$.
\subsubsection*{Conventional Fixed-Antenna System}
For comparison, we present the SR achieved by a conventional fixed-antenna system, where a single transmit antenna is placed at $[t_{\rm{f}},y_{\rm{t}},d]^{\mathsf{T}}$ and a single receive antenna is located at $[r_{\rm{f}},y_{\rm{r}},d]^{\mathsf{T}}$. The resulting SR is given by
\begin{align}\label{CASS_SR}
{\mathcal{R}}_{\rm{s}}^{\rm{f}}\triangleq\log_2\left(1+\frac{\overline{\gamma}_{\rm{s}}^{\rm{f}}}{d_{\rm{s}}^2+(t_{\rm{f}}-x_{\rm{c}})^2}\right),
\end{align}
where $\overline{\gamma}_{\rm{s}}^{\rm{f}}\triangleq\frac{PL\eta^2\alpha_{\rm{s}}}
{\sigma_{\rm{s}}^2(d_{\rm{r}}^2+(r_{\rm{f}}-x_{\rm{c}})^2)}$. Comparing \eqref{CASS_SR} with the SR under PASS, we note that the SR in PASS depends on the design of the activated or pinched locations $\mathbf{t}$, and that $\overline{\gamma}_{\rm{s}}^{\rm{f}}\leq \overline{\gamma}_{\rm{s}}$. These insights suggest that if optimized, PASS can deliver superior sensing performance over conventional fixed-antenna systems.

Note that both ${\mathcal{R}}_{\rm{c}}$ and ${\mathcal{R}}_{\rm{s}}$ depend on the pinching beamforming vector ${\mathbf{t}}$. However, finding an optimal $\mathbf{t}$ that maximizes both ${\mathcal{R}}_{\rm{c}}$ and ${\mathcal{R}}_{\rm{s}}$ simultaneously is a challenging task due to the conflicting nature of the objectives. To gain deeper insights into the system’s performance tradeoffs, we explore three representative design strategies: the \emph{C-C design}, the \emph{S-C design}, and the \emph{Pareto-optimal design}. Together, these approaches provide a full characterization of the achievable CR-SR region.
\section{Single-Pinch Case}\label{Section: Single-Pinch Case}
According to \cite{xu2024rate,ouyang2025array}, increasing the number of pinching antennas can significantly enhance the array gain achieved by a PASS, thereby improving both communication and sensing performance. However, realizing these gains requires the proper design of pinching beamforming, which is challenging due to the strongly coupled nature of the activated locations along the waveguide, as well as the dual phase shifts induced by signal propagation both within and outside the dielectric waveguide. To gain fundamental insights into the system design, we begin by analyzing the single-pinch case with only one active pinching antenna (i.e., $N=1$), where the CR and SR can be simplified as follows: ${\mathcal{R}}_{\rm{c}}=\log_2\left(1+\frac{\overline{\gamma}_{\rm{c}}}{d_{\rm{c}}^2+(t_1-x_{\rm{c}})^2}\right)$ and ${\mathcal{R}}_{\rm{s}}=\frac{1}{L}\log_2\left(1+\frac{\overline{\gamma}_{\rm{s}}}{d_{\rm{s}}^2+(t_1-x_{\rm{s}})^2}\right)$.
\subsection{Communications-Centric Design}\label{Section: Single-Pinch Case: Communications-Centric Design}
In the C-C design, the activated location $t_1$ is designed to maximize ${\mathcal{R}}_{\rm{c}}$, which satisfies
\begin{align}\label{SP_C_C_Optimal_Solution}
\argmax_{t_1}{\mathcal{R}}_{\rm{c}}=\argmax_{t_1}\frac{1}{{d_{\rm{c}}^2+(t_1-x_{\rm{c}})^2}}=x_{\rm{c}}.
\end{align}
Therefore, the CR and SR for the C-C design are given by
\begin{subequations}\label{Single_C_C_Rate}
\begin{align}
{\mathcal{R}}_{\rm{c}}&=\log_2\left(1+\frac{\overline{\gamma}_{\rm{c}}}{d_{\rm{c}}^2}\right),\\
{\mathcal{R}}_{\rm{s}}&=\frac{1}{L}\log_2\left(1+\frac{\overline{\gamma}_{\rm{s}}}
{d_{\rm{s}}^2+\Delta_x^2}\right),
\end{align}
\end{subequations}
respectively, where $\Delta_x\triangleq \lvert x_{\rm{c}}-x_{\rm{s}}\rvert$. 
\subsection{Sensing-Centric Design}
In the S-C design, the activated location $t_1$ is designed to maximize ${\mathcal{R}}_{\rm{s}}$, which satisfies
\begin{align}\label{SP_S_C_Optimal_Solution}
\argmax_{t_1}{\mathcal{R}}_{\rm{s}}=\argmax_{t_1}\frac{1}{{d_{\rm{s}}^2+(t_1-x_{\rm{s}})^2}}=x_{\rm{s}}.
\end{align}
Consequently, the CR and SR achieved by the S-C design are given by
\begin{subequations}\label{Single_S_C_Rate}
\begin{align}
{\mathcal{R}}_{\rm{c}}&=\log_2\left(1+\frac{\overline{\gamma}_{\rm{c}}}{d_{\rm{c}}^2+\Delta_x^2}\right),\\
{\mathcal{R}}_{\rm{s}}&=\frac{1}{L}\log_2\left(1+\frac{\overline{\gamma}_{\rm{s}}}
{d_{\rm{s}}^2}\right),
\end{align}
\end{subequations}
respectively. 
\vspace{-5pt}
\begin{remark}\label{SP_Position_Remark}
The results in \eqref{SP_C_C_Optimal_Solution} and \eqref{SP_S_C_Optimal_Solution} are intuitive, as they indicate that under the C-C and S-C designs, the optimal activated location corresponds to the point on the waveguide that is closest to the communication user or the sensing target, respectively. This placement minimizes the path loss in the corresponding communication or sensing channel, thereby maximizing the achievable CR or SR.
\end{remark}
\vspace{-5pt}
Comparing \eqref{Single_C_C_Rate} with \eqref{Single_S_C_Rate}, we observe that if $x_{\rm{c}}=x_{\rm{s}}$, i.e., the user and target are aligned along the waveguide axis, both CR and SR reach their respective maximum values simultaneously. However, as the offset $\lvert x_{\rm{c}}-x_{\rm{s}}\rvert$ increases, optimizing one functionality could lead to a degradation in the other. For example, when $\lvert x_{\rm{c}}-x_{\rm{s}}\rvert\rightarrow\infty$, the SR achieved under the C-C design approaches zero. 
\vspace{-5pt}
\begin{remark}\label{Subspace_Tradeoff_ISAC}
The above discussion implies that PASS-enabled ISAC systems achieve greater mutual benefits when the communication and sensing channels are spatially aligned.
\end{remark}
\vspace{-5pt}
\subsection{Pareto-Optimal Design}
To fully characterize the trade-off between communication and sensing performance, we investigate the Pareto boundary of the CR–SR region. This boundary includes all CR–SR pairs for which it is impossible to improve one rate without simultaneously decreasing the other. Each rate-tuple on the Pareto boundary can be obtained using the \emph{rate-profile} method \cite{zhang2010cooperative}, i.e., by solving the following optimization problem:
\begin{align}\label{SP_Pareto}
\max\nolimits_{t_1,\mathcal{R}}{\mathcal{R}},~{\rm{s.t.}}~{\mathcal{R}}_{\rm{c}}\geq\alpha{\mathcal{R}},~{\mathcal{R}}_{\rm{s}}\geq(1-\alpha){\mathcal{R}},
\end{align}
where $\alpha\in[0,1]$ is the \emph{rate-profile} parameter. The entire Pareto boundary can be traced by solving \eqref{SP_Pareto} across the range of $\alpha\in[0,1]$. Although this optimization problem is non-convex, it admits a closed-form solution, as provided below.
\vspace{-5pt}
\begin{theorem}\label{Theorem_SP_Pareto}
Given $\{\alpha,{\mathbf{u}}_{\rm{c}},{\mathbf{u}}_{\rm{s}}\}$, the optimal activated location $t_1$ of problem \eqref{SP_Pareto} is given by
\begin{align}
t_{\alpha}^{\star}=\left\{\begin{array}{ll}
                     x_{\rm{s}} & \frac{\log_2(1+\frac{\overline{\gamma}_{\rm{s}}}
{d_{\rm{s}}^2})}{L(1-\alpha)}<\frac{\log_2(1+\frac{\overline{\gamma}_{\rm{c}}}
{d_{\rm{c}}^2+\Delta_x^2})}{\alpha} \\
                     x_{\rm{c}} & \frac{\log_2(1+\frac{\overline{\gamma}_{\rm{s}}}
{d_{\rm{s}}^2+\Delta_x^2})}{L(1-\alpha)}>\frac{\log_2(1+\frac{\overline{\gamma}_{\rm{c}}}
{d_{\rm{c}}^2})}{\alpha} \\
                     x_{\rm{c}}+\beta_{\alpha}^{\star}\Delta_x & {\rm{Else}} 
                   \end{array}\right.,
\end{align}
where $\beta_{\alpha}^{\star}$ is the solution of $\beta$ to the equation $(1+\frac{\overline{\gamma}_{\rm{c}}}
{d_{\rm{c}}^2+\beta^2\Delta_x^2})^{L(1-\alpha)}=(1+\frac{\overline{\gamma}_{\rm{s}}}
{d_{\rm{s}}^2+(1-\beta)^2\Delta_x^2})^{\alpha}$ with $\beta\in[0,1]$.
\end{theorem}
\vspace{-5pt}
\begin{IEEEproof}
Refer to Appendix \ref{Proof_Theorem_SP_Pareto} for more details.
\end{IEEEproof}
\vspace{-5pt}
\begin{remark}
This result suggests that the optimal activated location lies along the line segment connecting the projections of the user and the target onto the waveguide axis, i.e., $t_{\alpha}^{\star}\in[x_{\rm{c}}, x_{\rm{s}}]$, which aligns with intuition. 
\end{remark} 
\vspace{-5pt}
For the case where $x_{\rm{s}}\leq x_{\rm{c}}$, a similar derivation yields the corresponding result.

Let ${\mathcal{R}}_{\rm{c}}^{t_{\alpha}^{\star}}$ and ${\mathcal{R}}_{\rm{s}}^{t_{\alpha}^{\star}}$ represent the CR and SR, respectively, achieved by setting $t_1=t_{\alpha}^{\star}$. The corresponding rate region consisting of all achievable rate pairs is defined as follows:
\begin{align}
{\mathcal{C}}_{\rm{S}}(t_{\alpha}^{\star})\triangleq\{({\mathcal{R}}_{\rm{c}},{\mathcal{R}}_{\rm{s}})\lvert
{\mathcal{R}}_{\rm{c}}\in[0,{\mathcal{R}}_{\rm{c}}^{t_{\alpha}^{\star}}],{\mathcal{R}}_{\rm{s}}\in[0,{\mathcal{R}}_{\rm{s}}^{t_{\alpha}^{\star}}]\}.
\end{align}
By flexibly adjusting the activated location $t_1$, any rate pair within the union set $\bigcup_{\alpha\in[0,1]}{\mathcal{C}}_{\rm{S}}(t_{\alpha}^{\star})$ can be achieved. Furthermore, applying the celebrated \emph{time-sharing} strategy \cite{el2011network,xiong2023fundamental} among different choices of $t_{\alpha}^{\star}$ (i.e., switching between different strategies with certain probabilities) leads to the \emph{complete} rate region of PASS, which is given by the convex hull of such a union set \cite{el2011network}:
\begin{align}\label{Rate_Region_SP}
{\mathcal{C}}_{\rm{p}}^{\rm{S}}\triangleq{\rm{Conv}}\left(\bigcup\nolimits_{\alpha\in[0,1]}{\mathcal{C}}_{\rm{S}}(t_{\alpha}^{\star})\right).
\end{align}

For comparison, we consider a conventional fixed-antenna system where the transmit and receive antennas are located at $[t_{\rm{f}},y_{\rm{t}},d]^{\mathsf{T}}$ and $[r_{\rm{f}},y_{\rm{r}},d]^{\mathsf{T}}$, respectively. The corresponding achievable rate region is given as follows:
\begin{align}
{\mathcal{C}}_{\rm{f}}\triangleq\{({\mathcal{R}}_{\rm{c}},{\mathcal{R}}_{\rm{s}})|{\mathcal{R}}_{\rm{c}}\in[0,{\mathcal{R}}_{\rm{c}}^{\rm{f}}],
{\mathcal{R}}_{\rm{s}}\in[0,{\mathcal{R}}_{\rm{s}}^{\rm{f}}]\},
\end{align}
where ${\mathcal{R}}_{\rm{c}}^{\rm{f}}$ and ${\mathcal{R}}_{\rm{s}}^{\rm{f}}$ are given in \eqref{CASS_CR} and \eqref{CASS_SR}, respectively. 

A rigorous mathematical comparison between ${\mathcal{R}}_{\rm{c}}^{\rm{f}}$ and ${\mathcal{R}}_{\rm{c}}^{t_{1}^{\star}}$ (the CR under the C-C design) is provided in \cite[Lemma 2]{ding2024flexible}, which states that it has
\begin{align}
{\mathcal{R}}_{\rm{c}}^{\rm{f}}>{\mathcal{R}}_{\rm{c}}^{t_{1}^{\star}}
\end{align}
when the user and target are uniformly distributed within a square region. Following a similar approach, it can also be shown that ${\mathcal{R}}_{\rm{s}}^{\rm{f}}>{\mathcal{R}}_{\rm{s}}^{t_{0}^{\star}}$, where ${\mathcal{R}}_{\rm{s}}^{t_{0}^{\star}}$ denotes the SR under the S-C design. 
\vspace{-5pt}
\begin{remark}
The above arguments imply that, under optimized pinching locations, the PASS system achieves higher communication and sensing performance than the conventional fixed-antenna system by reducing the large-scale path loss through large-scale antenna reconfiguration \cite{liu2025pinching}. 
\end{remark}
\vspace{-5pt}
In the following, we further compare the two rate regions ${\mathcal{C}}_{\rm{p}}^{\rm{S}}$ and ${\mathcal{C}}_{\rm{f}}$.
\vspace{-5pt}
\begin{theorem}\label{Theorem_SP_Capacity_Region}
The rate region achieved by the conventional fixed-antenna systems nests in the rate region achieved by PASS, namely ${\mathcal{C}}_{\rm{f}}\subseteq{\mathcal{C}}_{\rm{p}}^{\rm{S}}$.
\end{theorem}
\vspace{-5pt}
\begin{IEEEproof}
Please refer to Appendix \ref{Proof_Theorem_SP_Capacity_Region} for more details.
\end{IEEEproof}
\vspace{-5pt}
\begin{remark}
The above result suggests that the rate region achieved by the conventional fixed-location antenna system is completely contained within that of PASS.
\end{remark}
\vspace{-5pt}
\section{Multiple-Pinch Case}\label{Section: Multiple-Pinch Case}
We now consider the multiple-pinch case with $N>1$, where the DFSAC pinching beamforming must be carefully designed to achieve a favorable CR–SR tradeoff. As mentioned at the beginning of Section \ref{Section: Single-Pinch Case}, finding the optimal pinching beamformer is generally challenging. In the following, we address this challenge and aim to characterize the fundamental performance limits of multiple-pinch PASS-enabled ISAC.
\subsection{Communications-Centric Design}\label{Section: Multiple-Pinch Case: Communications-Centric Design}
In this case, the pinching beamformer ${\mathbf{t}}$ is optimized to maximize the CR as follows:
\begin{align}\label{C_C_Design_MP}
\argmax_{{\mathbf{t}}\in{\mathcal{P}}}\left\lvert\sum_{n=1}^{N}\frac{{{\rm{e}}^{-{\rm{j}}k_0\sqrt{d_{\rm{c}}^2+(t_n-x_{\rm{c}})^2}-{\rm{j}}{k_0t_n}n_{\rm{eff}}}}}
{\sqrt{d_{\rm{c}}^2+(t_n-x_{\rm{c}})^2}}\right\rvert^2\triangleq {\mathbf{t}}_{\rm{c}},
\end{align}
where ${\mathcal{P}}\triangleq\{\left.[t_1,\ldots,t_N]^{\mathsf{T}}\in{\mathbbmss{R}}^{N\times1}\right\rvert \lvert t_n-t_{n'}\rvert\geq\Delta,n\ne n'\}$ denotes the feasible set of ${\mathbf{t}}$, with $\Delta$ being the minimum spacing to mitigate mutual coupling \cite{ivrlavc2010toward}. Although problem \eqref{C_C_Design_MP} is non-convex and NP-hard, it can be optimally solved using the antenna location refinement method proposed in \cite{xu2024rate}. A comprehensive analysis of the achieved communication SNR $\gamma_{\rm{c}}$ can also be found in \cite{ouyang2025array}. Therefore, the detailed algorithm and theoretical analyses are omitted here due to space limitations. 
\subsection{Sensing-Centric Design}\label{Section: Multiple-Pinch Case: Sensing-Centric Design}
In this scenario, the pinching beamformer aims to maximize the SR as follows:
\begin{align}\label{S_C_Design_MP}
\argmax_{{\mathbf{t}}\in{\mathcal{P}}}\left\lvert\sum_{n=1}^{N}\frac{{{\rm{e}}^{-{\rm{j}}k_0\sqrt{d_{\rm{s}}^2+(t_n-x_{\rm{s}})^2}-{\rm{j}}{k_0t_n}n_{\rm{eff}}}}}
{\sqrt{d_{\rm{s}}^2+(t_n-x_{\rm{s}})^2}}\right\rvert^2\triangleq {\mathbf{t}}_{\rm{s}},
\end{align}
which can be solved by following steps similar to those in \cite{xu2024rate}. As discussed in \cite{xu2024rate,ouyang2025array}, under the C-C and S-C designs, the optimal pinching beamforming strategy aims to position the pinching antennas as close as possible to the communication user or sensing target, respectively, while satisfying the minimum inter-antenna distance constraint. This observation is consistent with the conclusion drawn in Remark \ref{SP_Position_Remark}, and further highlights the promising properties of PASS in mitigating path loss through flexible antenna placement.
\subsection{Pareto-Optimal Design}
We now investigate the Pareto-optimal design to characterize the CR-SR region by employing the \emph{rate-profile} method. This approach requires solving the following problem for different \emph{rate-profile} parameters $\alpha\in[0,1]$:
\begin{align}\label{MP_Pareto}
\max\nolimits_{{\mathbf{t}},\mathcal{R}}{\mathcal{R}},~{\rm{s.t.}}~{\mathcal{R}}_{\rm{c}}\geq\alpha{\mathcal{R}},~{\mathcal{R}}_{\rm{s}}\geq(1-\alpha){\mathcal{R}},~
{\mathbf{t}}\in{\mathcal{P}}.
\end{align}
\subsubsection{Rate Region}
Let ${\mathbf{t}}_{\alpha}^{\star}$ denote the optimal pinching beamforming vector for a given $\alpha\in[0,1]$, and let ${\mathcal{R}}_{\rm{c}}^{{\mathbf{t}}_{\alpha}^{\star}}$ and ${\mathcal{R}}_{\rm{s}}^{{\mathbf{t}}_{\alpha}^{\star}}$ denote the corresponding CR and SR, respectively. Following the derivations that lead to \eqref{Rate_Region_SP}, the rate region achieved by activating multiple pinching antennas is given by
\begin{align}\label{Rate_Region_MP}
{\mathcal{C}}_{\rm{p}}^{\rm{M}}\triangleq{\rm{Conv}}\left(\bigcup\nolimits_{\alpha\in[0,1]}{\mathcal{C}}_{\rm{M}}({\mathbf{t}}_{\alpha}^{\star})\right),
\end{align}
where
\begin{align}
{\mathcal{C}}_{\rm{M}}({\mathbf{t}}_{\alpha}^{\star})\triangleq\{({\mathcal{R}}_{\rm{c}},{\mathcal{R}}_{\rm{s}})\lvert
{\mathcal{R}}_{\rm{c}}\in[0,{\mathcal{R}}_{\rm{c}}^{{\mathbf{t}}_{\alpha}^{\star}}],{\mathcal{R}}_{\rm{s}}\in[0,{\mathcal{R}}_{\rm{s}}^{{\mathbf{t}}_{\alpha}^{\star}}]\}.
\end{align}

For $\alpha=1$ (the C-C design) and $\alpha=0$ (the S-C design), the optimal solutions to problem \eqref{MP_Pareto} have already been obtained in Sections \ref{Section: Multiple-Pinch Case: Communications-Centric Design} and \ref{Section: Multiple-Pinch Case: Sensing-Centric Design}, respectively. However, unlike the single-pinch case discussed in \eqref{SP_Pareto}, when $\alpha\in(0,1)$, the problem formulated in \eqref{MP_Pareto} does not admit a closed-form solution and typically requires an exhaustive search to determine the global optimum and the corresponding rate region ${\mathcal{C}}_{\rm{M}}({\mathbf{t}}_{\alpha}^{\star})$. To overcome this challenge and obtain more insights, we instead aim to develop both an \emph{inner bound} and an \emph{outer bound} on ${\mathcal{C}}_{\rm{p}}^{\rm{M}}$.
\subsubsection{Inner Bound}
In this section, we propose a suboptimal solution that yields an \emph{achievable inner bound} on the CR–SR region. 

To address the coupling among $\{t_n\}_{n=1}^{N}$, we adopt an element-wise alternating optimization framework. In this approach, each $t_n$ is optimized sequentially while keeping all other variables fixed. The subproblem for optimizing each $t_n$ can be formulated as follows:
\begin{subequations}\label{MP_Pareto_Sub}
\begin{align}
\max\nolimits_{t_n,\mathcal{R}}&{\mathcal{R}}\\{\rm{s.t.}}&~{\mathsf{R}}_{\rm{c}}^{(n)}(t_n)\geq\alpha{\mathcal{R}},{\mathsf{R}}_{\rm{s}}^{(n)}(t_n)\geq(1-\alpha){\mathcal{R}},\\
&~\lvert t_n-t_{n'} \rvert\geq\Delta,n\ne n',t_n\in[t_0,t_{\max}],\label{MP_Pareto_Sub_Cons1}
\end{align}
\end{subequations}
where ${\mathsf{R}}_{\rm{c}}^{(n)}(x)$ and ${\mathsf{R}}_{\rm{s}}^{(n)}(x)$ denote the CR and SR, respectively, when $t_n$ in $\mathbf{t}$ is set to $x$ and all other antenna locations remain fixed. Problem \eqref{MP_Pareto_Sub} is equivalent to the following:
\begin{align}\label{MP_Pareto_Sub_sub1}
\max_{t_n}\underbrace{\min\left\{\frac{1}{\alpha}{\mathsf{R}}_{\rm{c}}^{(n)}(t_n),\frac{1}{1-\alpha}{\mathsf{R}}_{\rm{s}}^{(n)}(t_n)\right\}}_{\triangleq~ {\mathsf{R}}_{\rm{IB}}^{(n)}(t_n)}~~{\rm{s.t.}}~\eqref{MP_Pareto_Sub_Cons1}.
\end{align}
Since this reduces to a single-variable optimization over a bounded interval, it can be efficiently solved by a low-complexity \emph{one-dimensional search}. Specifically, the feasible interval is first discretized into a uniform grid, and then the optimal $t_n$ is selected by evaluating the objective over the sampled points.

Let us discretize the interval $[t_0, t_{\max}]$ into a $Q$-point grid as follows:
\begin{align}
{\mathcal{Q}}\triangleq \left\{t_0,t_0+\frac{I_{\rm{P}}}{Q-1},t_0+\frac{2I_{\rm{P}}}{Q-1},\ldots,t_{\max}\right\},
\end{align}
where $I_{\rm{P}}\triangleq t_{\max}-t_0$. A near-optimal $t_n$, denoted by $\underline{t}_n^{\star}$, is then selected according to the following:
\begin{align}
\underline{t}_n^{\star}\triangleq\argmax\nolimits_{t_n\in {\mathcal{Q}}\setminus {\mathcal{Q}}_n}{\mathsf{R}}_{\rm{IB}}^{(n)}(t_n),
\end{align}
where the set ${\mathcal{Q}}_n$ includes all grid points that violate the minimum distance constraint, i.e.,
\begin{align}
{\mathcal{Q}}_n\triangleq \{x|x\in{\mathcal{Q}},\lvert x - t_{n'}\rvert<\Delta,n'\ne n\}.
\end{align}
The optimization proceeds by sequentially updating each $t_n$ across all $N$ antennas in an iterative manner until convergence. The full procedure is summarized in Algorithm \ref{Algorithm1}. The overall computational complexity of this algorithm scales as ${\mathcal{O}}(I_{\rm{iter}} NQ)$, where $I_{\rm{iter}}$ denotes the number of iterations until convergence.

\begin{algorithm}[!t]
\caption{Element-wise Algorithm for Solving \eqref{MP_Pareto}}
\label{Algorithm1}
\begin{algorithmic}[1]
\STATE initialize the optimization variables
\REPEAT 
  \FOR{$n\in\{1,\ldots,N\}$}
      \STATE update $t_n$ by solving problem \eqref{MP_Pareto_Sub_sub1} through one-dimensional search
    \ENDFOR
\UNTIL{the fractional decrease of the objective value of problem \eqref{MP_Pareto_Sub} falls below a predefined threshold}
\end{algorithmic}
\end{algorithm}

Let ${\underline{\mathbf{t}}}_{\alpha}^{\star}$ denote the optimized pinching beamformer corresponding to a given $\alpha\in(0,1)$, and $({\mathsf{R}}_{\rm{c}}^{\alpha},{\mathsf{R}}_{\rm{s}}^{\alpha})$ represent the achieved CR-SR pair. The corresponding rate region for a fixed $\alpha$ is given by
\begin{align}
{\mathcal{C}}_{\rm{M}}^{\rm{IB}}({\underline{\mathbf{t}}}_{\alpha}^{\star})\triangleq\{({\mathcal{R}}_{\rm{c}},{\mathcal{R}}_{\rm{s}})|{\mathcal{R}}_{\rm{c}}\in[0,{\mathsf{R}}_{\rm{c}}^{\alpha}],
{\mathcal{R}}_{\rm{s}}\in[0,{\mathsf{R}}_{\rm{s}}^{\alpha}]\}.
\end{align}
By further considering \emph{time sharing} among different ${\underline{\mathbf{t}}}_{\alpha}^{\star}$'s, the rate region of multiple-pinch PASS achieved by the element-wise algorithm is characterized by the following convex hull:
\begin{align}\label{Inner_Bound_ISAC}
{\mathcal{C}}_{\rm{p}}^{\rm{IB}}\triangleq{\rm{Conv}}\left({\mathcal{C}}_{\rm{M}}({\mathbf{t}}_{0}^{\star})\bigcup{\mathcal{C}}_{\rm{M}}({\mathbf{t}}_{1}^{\star})\bigcup\nolimits_{\alpha\in(0,1)}{\mathcal{C}}_{\rm{M}}^{\rm{IB}}({\underline{\mathbf{t}}}_{\alpha}^{\star})\right),
\end{align}
which satisfies ${\mathcal{C}}_{\rm{p}}^{\rm{IB}}\subseteq{\mathcal{C}}_{\rm{p}}^{\rm{M}}$. 

\begin{table*}[!t]
\centering
\caption{Summary of Main Results for Rate Region in PASS-Enabled ISAC}
\setlength{\abovecaptionskip}{0pt}
\resizebox{0.95\textwidth}{!}{
\begin{tabular}{|l|l|l|l|}
\hline
               & Communications-Centric                                                                                                 & Sensing-Centric                                                                                                        & Pareto-Optimal (Rate Region)                                                                                                                           \\ \hline
Single-Pinch   & ${\mathcal{C}}_{\rm{S}}(t_{1}^{\star})$ (\emph{closed-form}, $t_1=x_{\rm{c}}$)                                                & ${\mathcal{C}}_{\rm{S}}(t_{0}^{\star})$ (\emph{closed-form}, $t_1=x_{\rm{s}}$)                                                & ${\mathcal{C}}_{\rm{p}}^{\rm{S}}$ (\emph{closed-form}, $t_1=t_{\alpha}^{\star}$)                                                                              \\ \hline
Multiple-Pinch & ${\mathcal{C}}_{\rm{M}}({\mathbf{t}}_{1}^{\star})$ (\emph{method in \cite{xu2024rate}}, ${\mathbf{t}}={\mathbf{t}}_{\rm{c}}$) & ${\mathcal{C}}_{\rm{M}}({\mathbf{t}}_{0}^{\star})$ (\emph{method in \cite{xu2024rate}}, ${\mathbf{t}}={\mathbf{t}}_{\rm{s}}$) & \begin{tabular}[c]{@{}l@{}}${\mathcal{C}}_{\rm{p}}^{\rm{IB}}$ (\emph{rate-profile}); ${\mathcal{C}}_{\rm{p}}^{\rm{OB}}$ (\emph{closed-form})\\ ${\mathcal{C}}_{\rm{p}}^{\rm{IB}}\subseteq{\mathcal{C}}_{\rm{p}}\subseteq{\mathcal{C}}_{\rm{p}}^{\rm{OB}}$\end{tabular} \\ \hline
\end{tabular}}
\label{Table: PASS_Rate_Region}
\end{table*}

\subsubsection{Outer Bound} 
In this section, we aim to construct an \emph{outer bound} on the full rate region ${\mathcal{C}}_{\rm{p}}^{\rm{M}}$. By examining \eqref{Communication_SNR_Basic_Expression} and \eqref{Sensing_SNR_Basic_Expression}, we define the following set:
\begin{align}
{\mathcal{C}}_{\rm{OB}}^{\rm{I}}({\mathbf{t}})\triangleq\{({\mathsf{R}}_{\rm{c}},{\mathsf{R}}_{\rm{s}})\lvert
{\mathsf{R}}_{\rm{c}}\in[0,\overline{\mathsf{R}}_{\rm{c}}({\mathbf{t}})],
{\mathsf{R}}_{\rm{s}}\in[0,\overline{\mathsf{R}}_{\rm{s}}({\mathbf{t}})]\},
\end{align}
where ${\mathbf{t}}\in{\overline{\mathcal{P}}}\triangleq\{\left.[t_1,\ldots,t_N]^{\mathsf{T}}\right\rvert \lvert t_n\in[x_{\rm{c}},x_{\rm{s}}],n\in{\mathcal{N}}\}$, and
\begin{align}
\overline{\mathsf{R}}_{\rm{c}}({\mathbf{t}})&\triangleq\log_2\left(1+\overline{\gamma}_{\rm{c}}\left\lvert\sum_{n=1}^{N}\frac{1}
{\sqrt{d_{\rm{c}}^2+(t_n-x_{\rm{c}})^2}}\right\rvert^2\right),\label{Upper_Bound_CR_Initial}\\
\overline{\mathsf{R}}_{\rm{s}}({\mathbf{t}})&\triangleq\frac{1}{L}\log_2\left(1+\overline{\gamma}_{\rm{s}}\left\lvert\sum_{n=1}^{N}\frac{1}
{\sqrt{d_{\rm{s}}^2+(t_n-x_{\rm{s}})^2}}\right\rvert^2\right).\label{Upper_Bound_SR_Initial}
\end{align}
We define the convex hull over the union of ${\mathcal{C}}_{\rm{OB}}^{\rm{I}}({\mathbf{t}})$ for all feasible ${\mathbf{t}}\in{\overline{\mathcal{P}}}$ as follows:
\begin{align}\label{Set_Outer_Bound1}
{\mathcal{C}}_{\rm{OB}}^{(1)}\triangleq{\rm{Conv}}\left\{\bigcup\nolimits_{{\mathbf{t}}\in{\overline{\mathcal{P}}}}{\mathcal{C}}_{\rm{OB}}^{\rm{I}}({\mathbf{t}})\right\}.
\end{align}
This set is constructed by applying the following relaxations: \romannumeral1) \emph{ignoring} the minimum inter-antenna distance $\Delta$, \romannumeral2) \emph{neglecting} the dual phase shifts induced by signal propagation inside and outside the waveguide, and \romannumeral3) \emph{constraining} each activated location within the interval $[x_{\rm{c}},x_{\rm{s}}]$\footnote{The reason why the third relaxation yields an outer bound is that, by following steps similar to those in Appendix \ref{Proof_Lemma_SR}, the Pareto-optimal activated location of each pinching antenna---when both the minimum inter-antenna distance constraint (the first relaxation) and the dual phase shift (the second relaxation) are neglected---must satisfy $x_n\in[x_{\rm{c}},x_{\rm{s}}]$ for $n\in{\mathcal{N}}$.}. With these relaxations, for any achievable rate pair $({\mathcal{R}}_{\rm{c}},{\mathcal{R}}_{\rm{s}})\in{\mathcal{C}}_{\rm{p}}^{\rm{M}}$, there exists a corresponding point $({\mathsf{R}}_{\rm{c}},{\mathsf{R}}_{\rm{s}})\in{\mathcal{C}}_{\rm{OB}}^{(1)}$, which satisfies:
\begin{align}
{\mathcal{R}}_{\rm{c}}\leq {\mathsf{R}}_{\rm{c}},\quad {\mathcal{R}}_{\rm{s}}\leq{\mathsf{R}}_{\rm{s}}.
\end{align}
Thus, we conclude ${\mathcal{C}}_{\rm{p}}^{\rm{M}}\subseteq{\mathcal{C}}_{\rm{OB}}^{(1)}$. 

For clarity, we rewrite $t_n$ as follows:
\begin{align}
t_n=x_{\rm{c}}+\beta_n(x_{\rm{s}}-x_{\rm{c}})=x_{\rm{c}}+\beta_n\Delta_x,~n\in{\mathcal{N}},
\end{align} 
where $\beta_n\in[0,1]$ and $\Delta_x=\lvert x_{\rm{s}}-x_{\rm{c}}\rvert$. Substituting this into \eqref{Upper_Bound_CR_Initial} and \eqref{Upper_Bound_SR_Initial}, we have
\begin{align}
\overline{\mathsf{R}}_{\rm{c}}({\mathbf{t}})&=\log_2\left(1+\frac{\overline{\gamma}_{\rm{c}}}{\Delta_x^2}\left\lvert\sum_{n=1}^{N}\frac{1}
{\sqrt{{d_{\rm{c}}^2}/{\Delta_x^2}+\beta_n^2}}\right\rvert^2\right),\\
\overline{\mathsf{R}}_{\rm{s}}({\mathbf{t}})&=\frac{1}{L}\log_2\left(1+\frac{\overline{\gamma}_{\rm{s}}}{\Delta_x^2}\left\lvert\sum_{n=1}^{N}\frac{1}
{\sqrt{{d_{\rm{s}}^2}/{\Delta_x^2}+(1-\beta_n)^2}}\right\rvert^2\right).
\end{align}

A full characterization of ${\mathcal{C}}_{\rm{OB}}^{(1)}$ requires an exhaustive search over the \emph{unit hypercube} $[0,1]^{N}$, which is mathematically intractable. To overcome this challenge, we apply the \emph{Cauchy–Schwarz inequality} to derive the following relaxation:
\begin{align}\label{Upper_Bound_Cauchy–Schwarz}
\left\lvert\sum_{n=1}^{N}\frac{1}
{\sqrt{{d_{\rm{c}}^2}/{\Delta_x^2}+\beta_n^2}}\right\rvert^2\leq \sum_{n=1}^{N}\frac{N}
{{{d_{\rm{c}}^2}/{\Delta_x^2}+\beta_n^2}}.
\end{align}
Furthermore, we observe that $\frac{1}
{{{d_{\rm{c}}^2}/{\Delta_x^2}+\beta_n^2}}$ is a strictly convex function of $\beta_n^2$ over $[0,1]$. This allows the application of \emph{Karamata's inequality} (also known as the \emph{majorization inequality} \cite{kadelburg2005inequalities}) to further simplify the analysis.
\vspace{-5pt}
\begin{lemma}\label{Extreme_Value_Lemma}
Given ${\bm\beta}\triangleq[\beta_1,\ldots,\beta_N]^{\mathsf{T}}\in[0,1]^N$, define $S_{\bm\beta}\triangleq\sum_{n=1}^{N}\beta_n^2\in[0,N]$ and ${\underline{S}}_{\bm\beta}\triangleq \lfloor{S}_{\bm\beta}\rfloor$. Then, for $\rho\ne 0$, it has
\begin{align}
\sum_{n=1}^{N}\frac{1}
{{\rho^2+\beta_n^2}}&\leq 
\frac{{\underline{S}}_{\bm\beta}}{\rho^2+1}+\frac{{\mathbbmss{1}}_{S_{\bm\beta}\in{\mathbbmss{Z}}}}{\rho^2+{S}_{\bm\beta}-{\underline{S}}_{\bm\beta}}
+\frac{N-{\underline{S}}_{\bm\beta}-{\mathbbmss{1}}_{S_{\bm\beta}\in{\mathbbmss{Z}}}}{\rho^2}\nonumber\\
&\triangleq f_{\rm{UB}}(\rho^2,S_{\bm\beta}),\label{Extreme_Value_Lemma_Most_Important}
\end{align}
where ${\mathbbmss{1}}_{S_{\bm\beta}\in{\mathbbmss{Z}}}$ is the indicator function that equals $1$ if $S_{\bm\beta}$ is an integer and $0$ otherwise.
\end{lemma}
\vspace{-5pt}
\begin{IEEEproof}
Refer to Appendix \ref{Proof_Extreme_Value_Lemma} for more details.
\end{IEEEproof}
Based on \eqref{Upper_Bound_Cauchy–Schwarz} and Lemma \ref{Extreme_Value_Lemma}, we can upper bound $\overline{\mathsf{R}}_{\rm{c}}({\mathbf{t}})$ as follows:
\begin{align}\label{Upper_Bound_CR_Karamata}
\overline{\mathsf{R}}_{\rm{c}}({\mathbf{t}})\leq 
\log_2\left(1+\frac{\overline{\gamma}_{\rm{c}}N}{\Delta_x^2}f_{\rm{UB}}\left(\frac{d_{\rm{c}}^2}{\Delta_x^2},S_{\bm\beta}\right)\right),
\end{align}
and similarly,
\begin{equation}\label{Upper_Bound_SR_Karamata}
\overline{\mathsf{R}}_{\rm{s}}({\mathbf{t}})\leq
\frac{1}{L}\log_2\left(1+\frac{\overline{\gamma}_{\rm{s}}N}{\Delta_x^2}f_{\rm{UB}}\left(\frac{d_{\rm{s}}^2}{\Delta_x^2},\overline{S}_{{\bm\beta}}\right)\right),
\end{equation}
where $\overline{S}_{{\bm\beta}}\triangleq\sum_{n=1}^{N}(1-\beta_n)^2\in[0,N]$.

Referring to \eqref{Extreme_Value_Lemma_Most_Important}, we observe that the function $f_{\rm{UB}}(\rho^2,x)$ is monotonically decreasing with respect to $x$ for $x\in[0,N]$. By applying the \emph{Cauchy–Schwarz inequality}, we obtain the following bound:
\begin{align}\label{CS_Inequality_Result_Then}
S_{\bm\beta}=\sum_{n=1}^{N}\beta_n^2\geq \frac{1}{N} Z_{\bm\beta}^2,
\end{align}
where $Z_{\bm\beta}\triangleq\sum_{n=1}^{N}\beta_n\in[0,N]$. Combining \eqref{CS_Inequality_Result_Then} with \eqref{Upper_Bound_CR_Karamata} yields the following upper bound of $\overline{\mathsf{R}}_{\rm{c}}({\mathbf{t}})$:
\begin{align}\label{Upper_Bound_CR_Karamata_Cauchy}
\overline{\mathsf{R}}_{\rm{c}}({\mathbf{t}})\leq 
\log_2\left(1+\frac{\overline{\gamma}_{\rm{c}}N}{\Delta_x^2}f_{\rm{UB}}\left(\frac{d_{\rm{c}}^2}{\Delta_x^2},\frac{Z_{\bm\beta}^2}{N} \right)\right)\triangleq \hat{\mathsf{R}}_{\rm{c}}(Z_{\bm\beta}),
\end{align}
and similarly,
\begin{equation}\label{Upper_Bound_SR_Karamata_Cauchy}
\begin{split}
\overline{\mathsf{R}}_{\rm{s}}({\mathbf{t}})&\leq
\frac{1}{L}\log_2\left(1+\frac{\overline{\gamma}_{\rm{s}}N}{\Delta_x^2}f_{\rm{UB}}\left(\frac{d_{\rm{s}}^2}{\Delta_x^2},\frac{(N-Z_{\bm\beta})^2}{N}\right)\right)\\
&\triangleq \hat{\mathsf{R}}_{\rm{s}}(Z_{\bm\beta}).
\end{split}
\end{equation}
Thus, we can obtain the upper bounds for $\overline{\mathsf{R}}_{\rm{c}}({\mathbf{t}})$ and $\overline{\mathsf{R}}_{\rm{s}}({\mathbf{t}})$, which now depend solely on $Z_{\bm\beta}$, viz. the sum of $\{\beta_n\}_{n=1}^{N}$. For ease of clarification, we define the following region associated with each $Z_{\bm\beta}\in[0,N]$:
\begin{align}
{\mathcal{C}}_{Z_{\bm\beta}}\triangleq\{({\mathsf{R}}_{\rm{c}},{\mathsf{R}}_{\rm{s}})|{\mathsf{R}}_{\rm{c}}\in[0,\hat{\mathsf{R}}_{\rm{c}}(Z_{\bm\beta})],
{\mathsf{R}}_{\rm{s}}\in[0,\hat{\mathsf{R}}_{\rm{s}}(Z_{\bm\beta})]\}.
\end{align}
By allowing $Z_{\bm\beta}$ to vary from $0$ to $N$, we define the outer bound region as the convex hull of the union of all such regions:
\begin{align}\label{Outer_Bound_ISAC}
{\mathcal{C}}_{\rm{p}}^{\rm{OB}}\triangleq {\rm{Conv}}\left(\bigcup\nolimits_{Z_{\bm\beta}\in[0,N]}{\mathcal{C}}_{Z_{\bm\beta}}\right).
\end{align}
From \eqref{Upper_Bound_CR_Karamata_Cauchy} and \eqref{Upper_Bound_SR_Karamata_Cauchy}, it follows that for any point $({\mathsf{R}}_{\rm{c}},{\mathsf{R}}_{\rm{s}})\in{\mathcal{C}}_{\rm{OB}}^{(1)}$, there always exists a point $({\mathsf{R}}_{\rm{c}}^{\rm{p}},{\mathsf{R}}_{\rm{s}}^{\rm{p}})\in{\mathcal{C}}_{\rm{p}}^{\rm{OB}}$ such that:
\begin{align}
{\mathsf{R}}_{\rm{c}}\leq {\mathsf{R}}_{\rm{c}}^{\rm{p}},\quad {\mathsf{R}}_{\rm{s}}\leq {\mathsf{R}}_{\rm{s}}^{\rm{p}}.
\end{align}
This inclusion implies ${\mathcal{C}}_{\rm{OB}}^{(1)}\subseteq{\mathcal{C}}_{\rm{p}}^{\rm{OB}}$. Taken together, we conclude
\begin{align}
{\mathcal{C}}_{\rm{p}}^{\rm{M}}\subseteq{\mathcal{C}}_{\rm{OB}}^{(1)}\subseteq{\mathcal{C}}_{\rm{p}}^{\rm{OB}}.
\end{align}

The above upper bound applies to the case where $\Delta_x\ne0$. For the special case of $\Delta_x=0$ or when $x_{\rm{c}}=x_{\rm{s}}$, closed-form upper bounds for both the CR and SR are available in \cite{ouyang2025array}, where their tightness has also been validated through numerical simulations. Due to page limitations, we omit the detailed derivations and refer the interested reader to \cite{ouyang2025array} for further information.

Thus, we have successfully established both an \emph{inner bound} (\eqref{Inner_Bound_ISAC}) and an \emph{outer bound} (\eqref{Outer_Bound_ISAC}) for the rate region achieved by PASS with multiple pinching antennas: 
\begin{align}
{\mathcal{C}}_{\rm{p}}^{\rm{IB}}\subseteq{\mathcal{C}}_{\rm{p}}^{\rm{M}}\subseteq{\mathcal{C}}_{\rm{p}}^{\rm{OB}}.
\end{align}

For ease of reference, we summarize in Table \ref{Table: PASS_Rate_Region} our main results on the rate region characterization.

\begin{figure}[!t]
\centering
\includegraphics[width=0.45\textwidth]{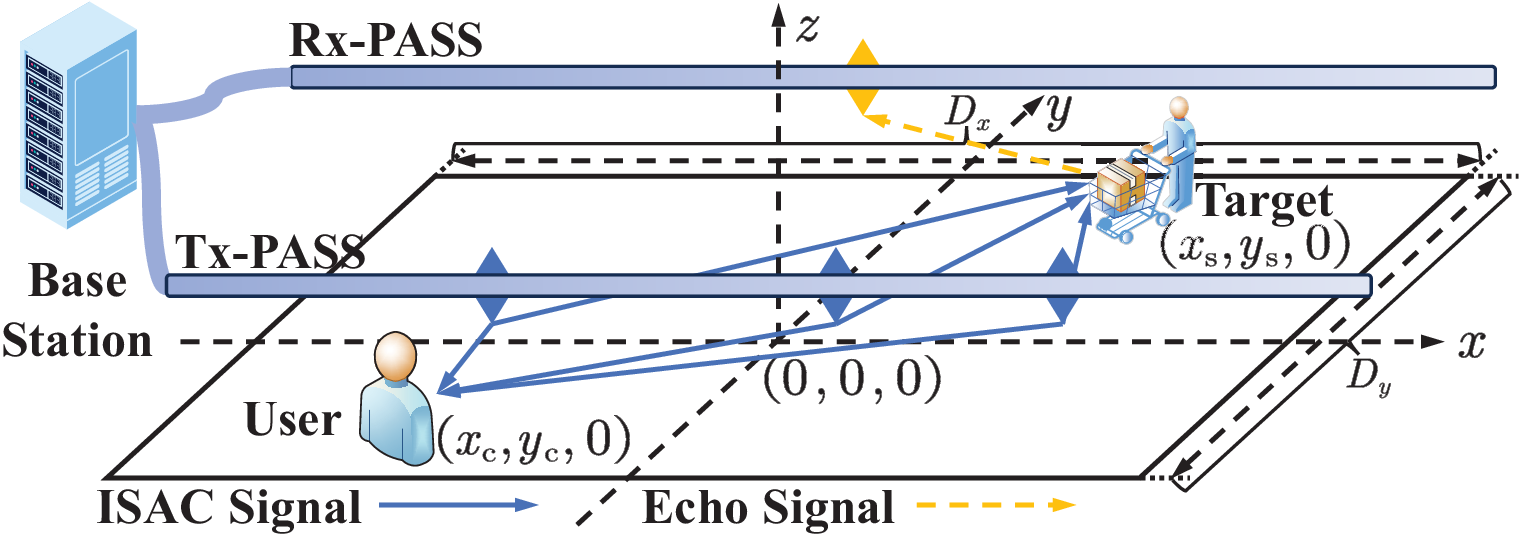}
\caption{Illustration of the simulation setup.}
\label{Figure_simulation}
\vspace{-10pt}
\end{figure}

\section{Numerical Results}\label{Section_Numerical_Results}
In this section, we present numerical results to validate our analytical findings. Unless otherwise specified, the communication user and sensing target are uniformly distributed within a rectangular region centered at the origin, with side lengths $D_x=20$ m and $D_y=8$ m along the $x$- and $y$-axes, respectively, as illustrated in {\figurename} {\ref{Figure_simulation}}. The transmit and receive waveguides are deployed at a height of $d=3$ m, with their positions fixed along the $x$-axis at $y_{\rm{t}}=-2$ m and $y_{\rm{r}}=2$ m, respectively. The $x$-coordinate of the feed point is set to $t_0=r_0=-\frac{D_x}{2}$ for the single-pinch case, and $t_0=r_0=-\frac{D_x}{2}-1$ m for the multiple-pinch case. The maximum deployment range is set to $t_{\max}=\frac{D_x}{2}$ for the single-pinch case, and $t_{\max}=\frac{D_x}{2}+1$ m for the multiple-pinch case. The default simulation parameters are as follows: carrier frequency $f_{\rm{c}}=28$ GHz, effective refractive index $n_{\rm{eff}}=1.4$, ISAC frame length $L=5$, average reflection strength $\alpha_{\rm{s}}=10$, one-dimensional search precision $Q=10^4$, minimum inter-antenna distance $\Delta=\frac{\lambda}{2}$, transmit power $P=10$ dBm, and noise power $\sigma_{\rm{c}}^2=\sigma_{\rm{s}}^2=-114$ dBm. 

Two PASS configurations are considered: \romannumeral1) Case {\uppercase\expandafter{\romannumeral1}}: an ideal waveguide with no in-waveguide propagation loss and \romannumeral2) Case {\uppercase\expandafter{\romannumeral2}}: a practical waveguide with an in-waveguide propagation loss factor of $0.08$ dB/m \cite{ding2024flexible,wang2024antenna,ouyang2025array}. For Case {\uppercase\expandafter{\romannumeral2}}, the rate region is computed using the activated locations optimized under Case {\uppercase\expandafter{\romannumeral1}}, with in-waveguide propagation loss subsequently incorporated. For comparison, we also include results for a conventional fixed-antenna system with a single transmit antenna at $[0,y_{\rm{t}},d]^{\mathsf{T}}$ and a single receive antenna at $[0,y_{\rm{r}},d]^{\mathsf{T}}$. All results are averaged over $10^3$ independent channel realizations.

\subsection{Single-Pinch Case}
\subsubsection{Achievable CR and SR}
\begin{figure}[!t]
\centering
    \subfigure[CR vs. $D_x$.]
    {
        \includegraphics[width=0.4\textwidth]{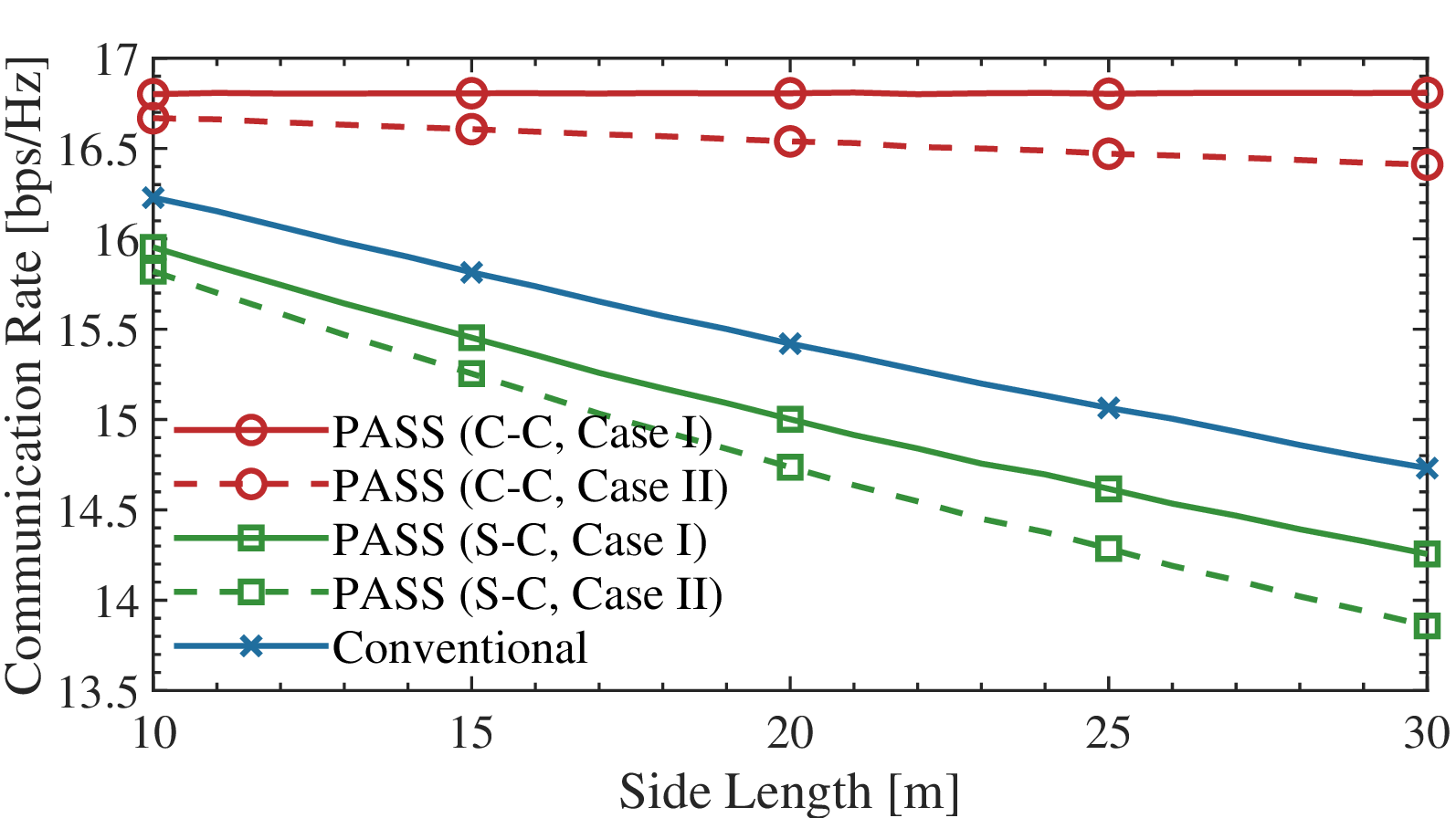}
	   \label{Figure_SP_CR_Range}
    }
   \subfigure[SR vs. $D_x$.]
    {
        \includegraphics[width=0.4\textwidth]{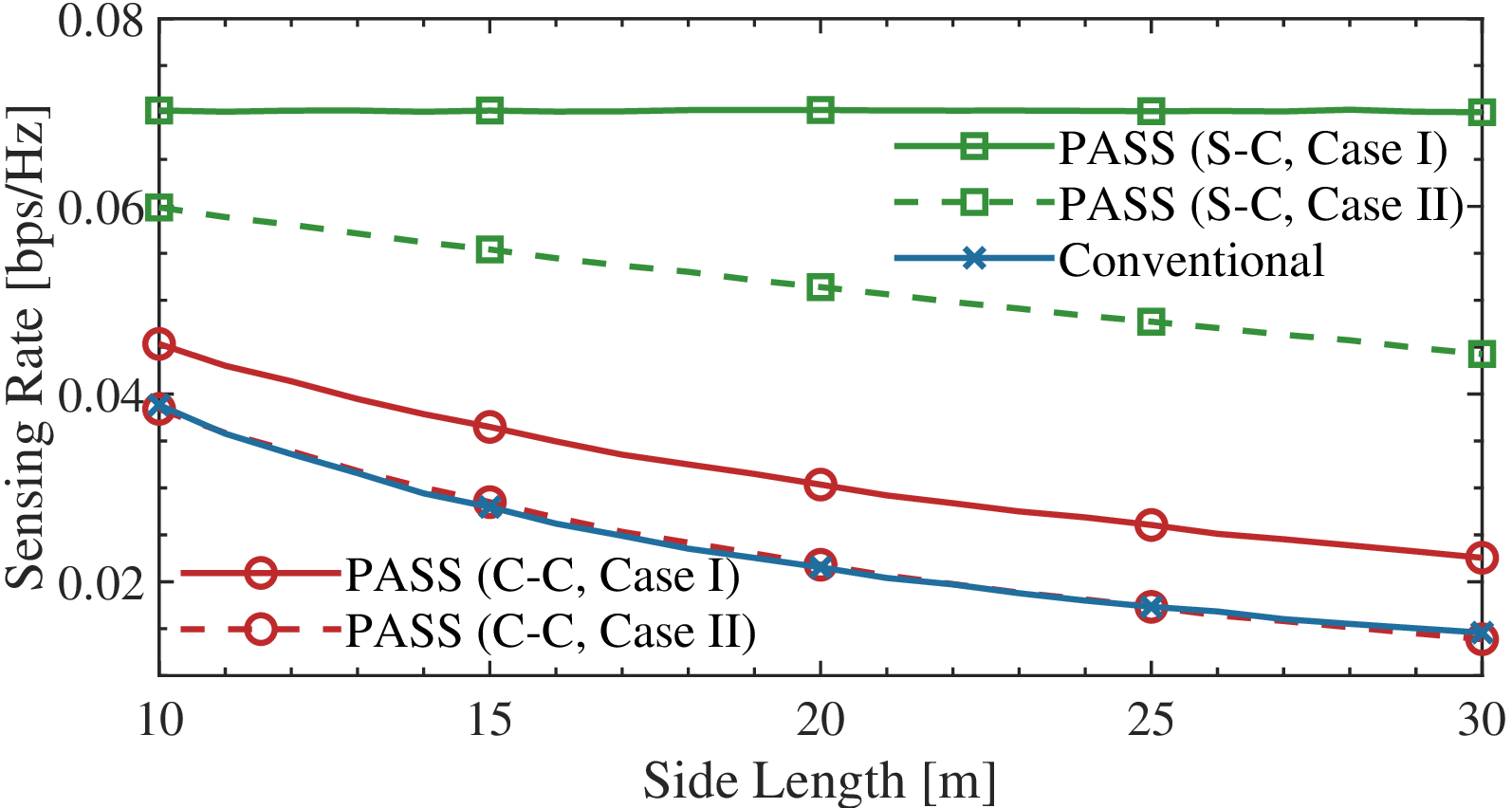}
	   \label{Figure_SP_SR_Range}
    }
\caption{Achievable CR and SR of single-pinch PASS.}
\label{Figure: CR_SR_Single_Pinch_PASS}
\vspace{-10pt}
\end{figure}

{\figurename} {\ref{Figure: CR_SR_Single_Pinch_PASS}} compares the CR and SR achieved by the conventional fixed-antenna system and the proposed single-pinch PASS architecture under both C-C and S-C designs. 

{\figurename} {\ref{Figure_SP_CR_Range}} illustrates the achievable CR as a function of the side length $D_x$. It is observed that the C-C design for PASS demonstrates superior CR performance compared with the S-C design, irrespective of in-waveguide propagation loss. Furthermore, the in-waveguide propagation loss causes only marginal CR degradation for both C-C and S-C designs, which is consistent with findings in \cite{ding2024flexible,ouyang2025array,wang2024antenna}. Consequently, the subsequent discussion focuses on Case {\uppercase\expandafter{\romannumeral1}} (i.e., the \emph{ideal} waveguide scenario without in-waveguide propagation loss). 

In the C-C design, the antenna is activated at the projection of the communication user's location onto the transmit waveguide axis. This placement ensures that the user experiences the same path loss while moving along the waveguide axis, which results in a CR that remains invariant with respect to $D_x$. By contrast, both the fixed-antenna system and the S-C PASS exhibit a monotonic decline in CR as $D_{x}$ increases, due to the growing average distance between the user and either the waveguide center (fixed-antenna) or the S-C activated location. As anticipated, the C-C PASS achieves a higher CR than the fixed-antenna benchmark. However, the S-C PASS underperforms due to inherent spatial constraints. Specifically, for the uniformly distributed user and target, the distance between them along the waveguide axis, $\Delta_x= \lvert x_{\rm{c}}-x_{\rm{s}}\rvert$, follows a \emph{triangular distribution} over $[0,D_x]$. In comparison, the distance from the user to the fixed transmit antenna (i.e., the waveguide center), $\lvert x_{\rm{c}}\rvert$, follows a \emph{uniform distribution} over $[0,\frac{D_x}{2}]$. The triangular distribution in the S-C PASS case results in a \emph{larger average transmission distance} compared to the conventional-antenna system, which leads to its reduced CR performance.

{\figurename} {\ref{Figure_SP_SR_Range}} depicts the SR as a function of the side length $D_x$. It is observed that in-waveguide propagation loss has a more significant impact on SR than the CR shown in {\figurename} {\ref{Figure_SP_CR_Range}}. This is because SR is affected by signal attenuation in both the transmit and receive waveguides, whereas CR is solely influenced by losses in the transmit waveguide. Despite this dual-waveguide attenuation, the S-C design still enables PASS to achieve a higher SR than the conventional fixed-antenna system, which confirms that the impact of in-waveguide propagation loss is insignificant. Another important observation from {\figurename} {\ref{Figure_SP_SR_Range}} is that the C-C design either \emph{outperforms or closely matches} the SR of the conventional system. This is in contrast to the CR behavior in {\figurename} {\ref{Figure_SP_CR_Range}}, where the S-C PASS \emph{underperforms} compared to the conventional-antenna system. This is explained as follows. In the sensing process, the \emph{receive-side activated location} $r$ is aligned with the target's true location $x_{\rm{s}}$. This alignment minimizes performance degradation caused by the mismatch between $x_{\rm{s}}$ and the \emph{transmit-side activated position} $x_{\rm{c}}$ under the C-C design, even when in-waveguide propagation loss is considered.

\begin{figure}[!t]
\centering
\includegraphics[width=0.4\textwidth]{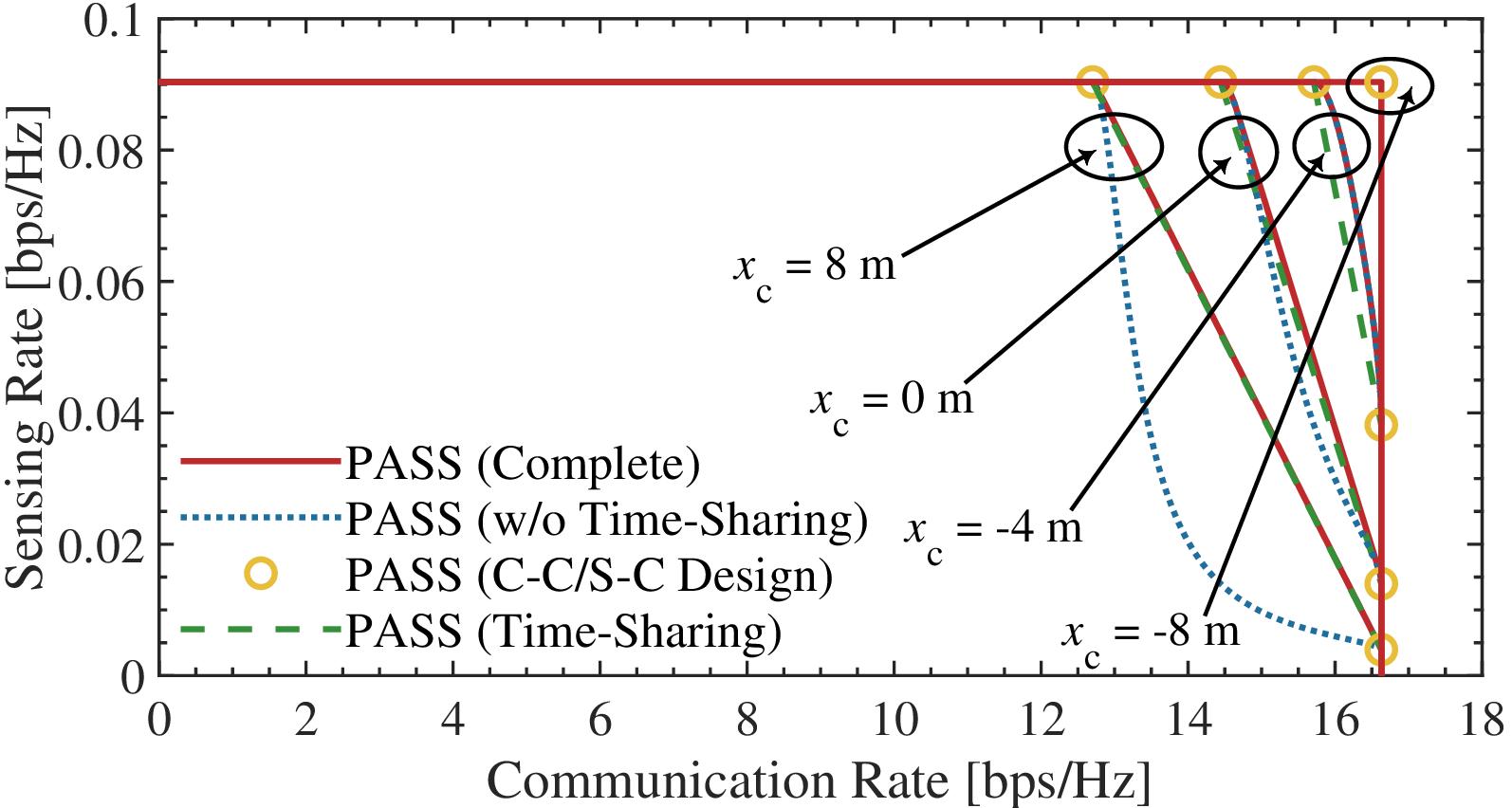}
\caption{Instantaneous rate region of single-pinch PASS (Case {\uppercase\expandafter{\romannumeral1}}). $y_{\rm{c}}=1$ m, $y_{\rm{s}}=-1$ m, and $x_{\rm{s}}=-8$ m.}
\label{Figure_SP_Rate_Range}
\vspace{-10pt}
\end{figure}

\subsubsection{Rate Region}
{\figurename} {\ref{Figure_SP_Rate_Range}} illustrates the \emph{instantaneous} rate region achieved by the single-pinch PASS under various values of $x_{\rm{c}}$, while fixing $y_{\rm{c}}=1$ m, $y_{\rm{s}}=-1$ m, and $x_{\rm{s}}=-8$ m. For comparison, we plot: \romannumeral1) the \emph{complete} rate region computed from the rate-profile method with time sharing (i.e., the region defined in \eqref{Rate_Region_SP}), \romannumeral2) the \emph{achievable} region using the rate-profile method \emph{without} time sharing (i.e., $\bigcup\nolimits_{\alpha\in[0,1]}{\mathcal{C}}_{\rm{S}}(t_{\alpha}^{\star})$), and \romannumeral3) the rate region obtained by simply \emph{time sharing} between the C-C and S-C designs. For brevity, only Case {\uppercase\expandafter{\romannumeral1}} is considered. As shown in this graph, when $x_{\rm{c}}$ deviates significantly from $x_{\rm{s}}$, e.g., $x_{\rm{c}}=8$ m, a simple time-sharing scheme between C-C and S-C designs is sufficient to attain the complete rate region. However, as $x_{\rm{c}}$ moves closer to $x_{\rm{s}}$, fully achieving the optimal rate region requires employing both the rate-profile method and time sharing, as defined by \eqref{Rate_Region_SP}. Additionally, we observe that as $x_{\rm{c}}$ gradually converges toward $x_{\rm{s}}$, the shape of the achievable rate region becomes increasingly \emph{rectangular}. This transformation indicates a diminishing trade-off between CR and SR, as the communication user and the sensing target become spatially aligned along the transmit waveguide axis. This observation is consistent with our earlier discussion in Remark \ref{Subspace_Tradeoff_ISAC}.

\begin{figure}[!t]
\centering
\includegraphics[width=0.4\textwidth]{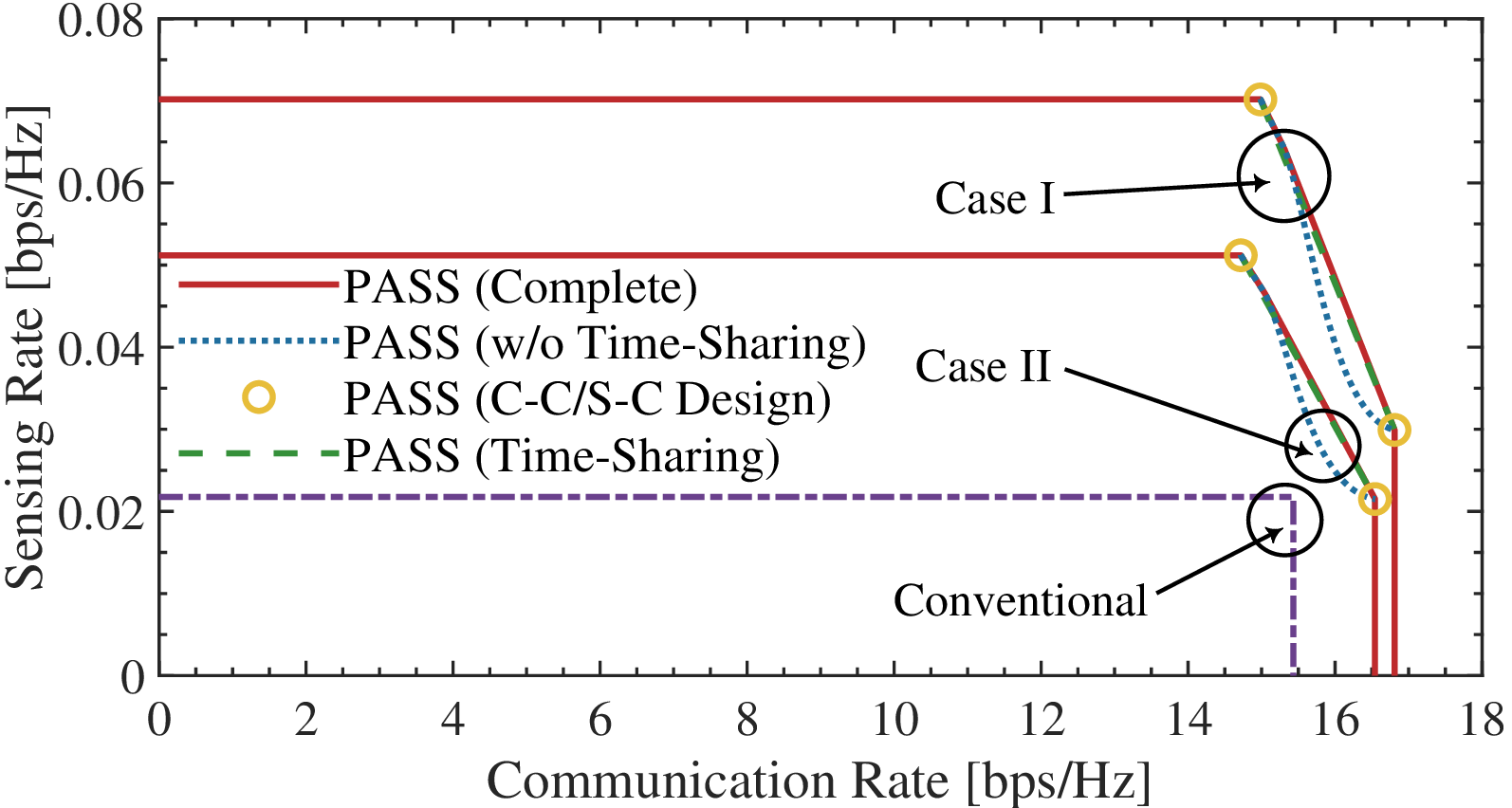}
\caption{Average rate region of single-pinch PASS.}
\label{Figure_SP_Average_Rate_Range}
\vspace{-10pt}
\end{figure}

By further considering the uniform distribution of both the communication user and the sensing target, {\figurename} {\ref{Figure_SP_Average_Rate_Range}} compares the \emph{average} rate region achieved by the single-pinch PASS and that of the conventional fixed-antenna system. To enable a comprehensive comparison, we also present the rate regions under the impact of in-waveguide propagation loss. As can be seen from this graph, on average, employing a time-sharing strategy between the S-C and C-C designs enables the single-pinch PASS to closely approach the complete rate region. Additionally, although in-waveguide propagation loss narrows the achievable rate region, PASS still outperforms the conventional-antenna system by achieving a larger rate region, which validates the correctness of Theorem \ref{Theorem_SP_Capacity_Region}. Furthermore, it is found that the in-waveguide loss has a more pronounced impact on SR compared to CR, which aligns with the trends previously observed in {\figurename} {\ref{Figure: CR_SR_Single_Pinch_PASS}}.

\begin{figure}[!t]
\centering
    \subfigure[CR vs. $N$.]
    {
        \includegraphics[width=0.4\textwidth]{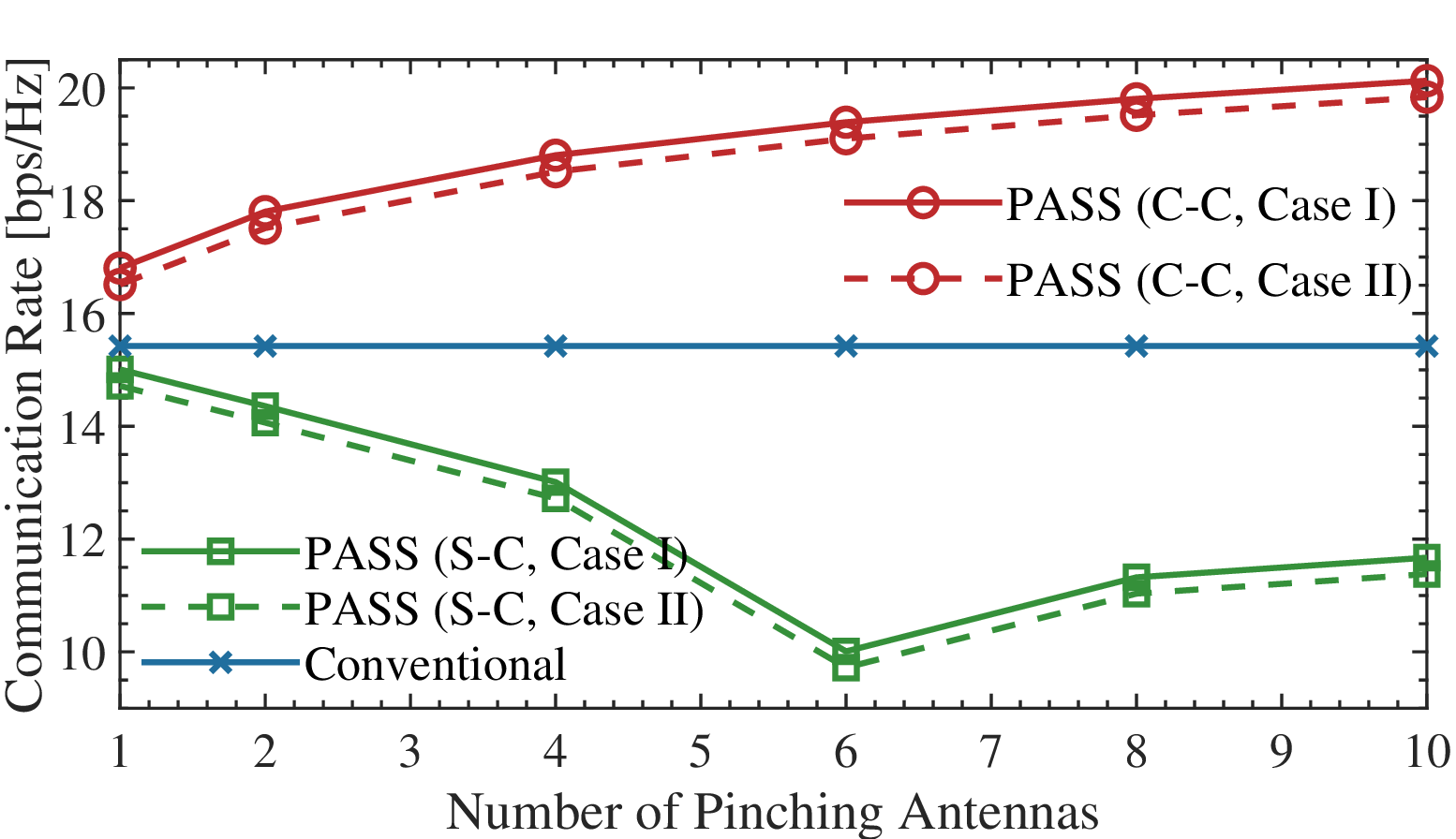}
	   \label{Figure_MP_CR_Antenna}
    }
   \subfigure[SR vs. $N$.]
    {
        \includegraphics[width=0.4\textwidth]{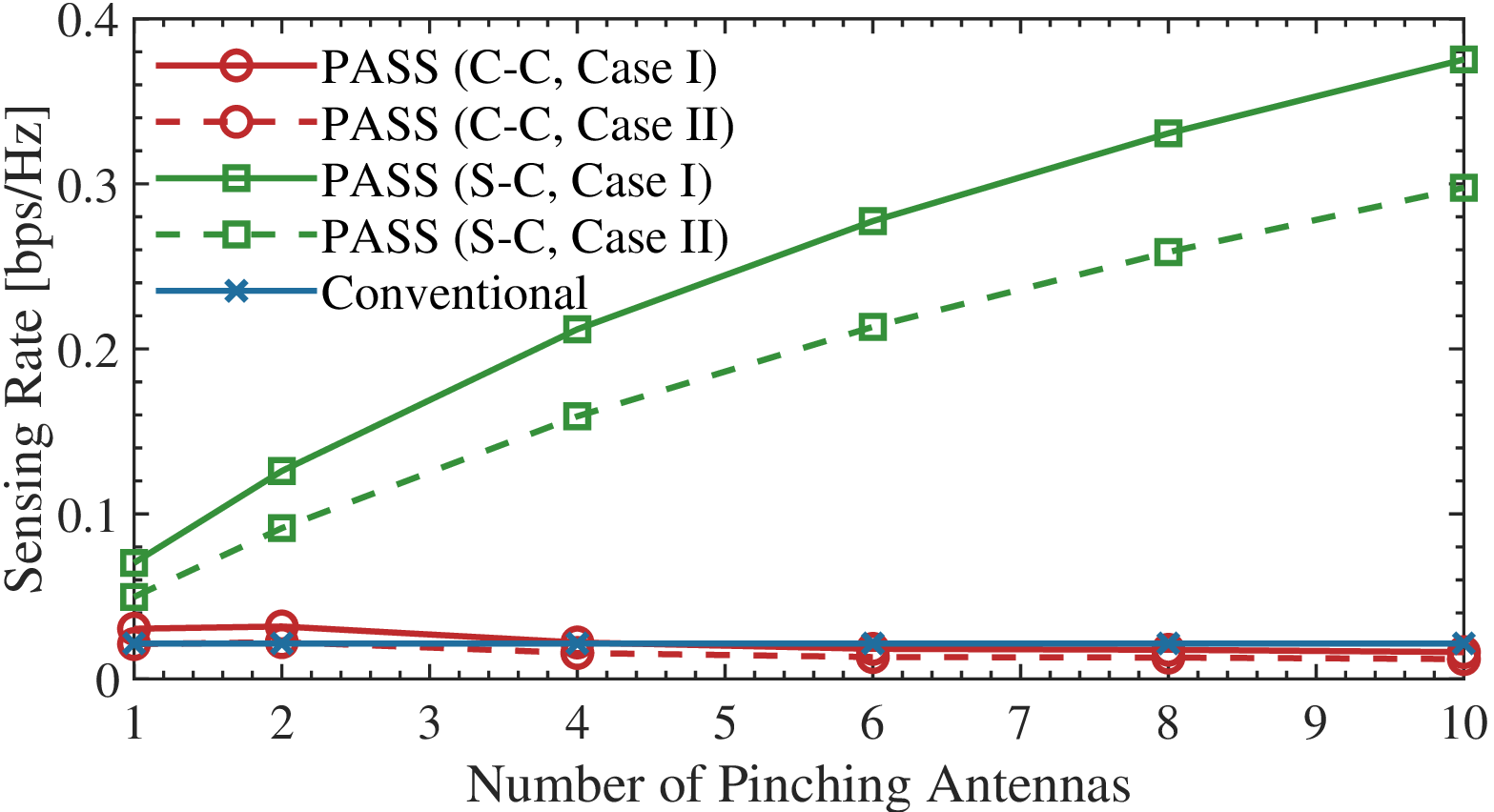}
	   \label{Figure_MP_SR_Antenna}
    }
\caption{Achievable CR and SR of multiple-pinch PASS.}
\label{Figure: CR_SR_Multiple_Pinch_PASS}
\vspace{-10pt}
\end{figure}

\subsection{Multiple-Pinch Case}
\subsubsection{Achievable CR and SR}
In this section, the multiple-pinch configuration is focused on, where the achievable CR and SR are evaluated as functions of the number of pinching antennas, $N$, as shown in {\figurename} {\ref{Figure_MP_CR_Antenna}} and {\figurename} {\ref{Figure_MP_SR_Antenna}}, respectively.

From {\figurename} {\ref{Figure_MP_CR_Antenna}}, it is observed that under the C-C design, the CR achieved by PASS increases monotonically with $N$ for both Case {\uppercase\expandafter{\romannumeral1}} and Case {\uppercase\expandafter{\romannumeral2}}. Moreover, the presence of in-waveguide propagation loss results in only a negligible degradation in CR, which is consistent with the findings in \cite{ouyang2025array}. Notably, regardless of whether in-waveguide propagation loss is considered, the C-C PASS outperforms the conventional fixed-antenna system in terms of CR. In contrast, the S-C design yields a lower CR than the conventional system, which is attributed to the inherent mismatch between the sensing-optimal pinching beamformer and the communication user's spatial channel. {\figurename} {\ref{Figure_MP_SR_Antenna}} focuses on the SR performance. It is shown that under the S-C design, the SR increases with $N$ for both Case {\uppercase\expandafter{\romannumeral1}} and Case {\uppercase\expandafter{\romannumeral2}}. In both cases, the S-C PASS achieves superior SR compared to the conventional fixed-antenna setup. However, in contrast to CR, the SR is more severely impacted by in-waveguide propagation loss due to signal attenuation occurring in both the transmit and receive waveguides. As for the C-C design, the resulting SR either exceeds or closely matches that of the conventional system, which confirms the observations made earlier in {\figurename} {\ref{Figure_SP_SR_Range}}.

\begin{figure}[!t]
\centering
    \subfigure[$x_{\rm{c}}=8$ m.]
    {
        \includegraphics[width=0.22\textwidth]{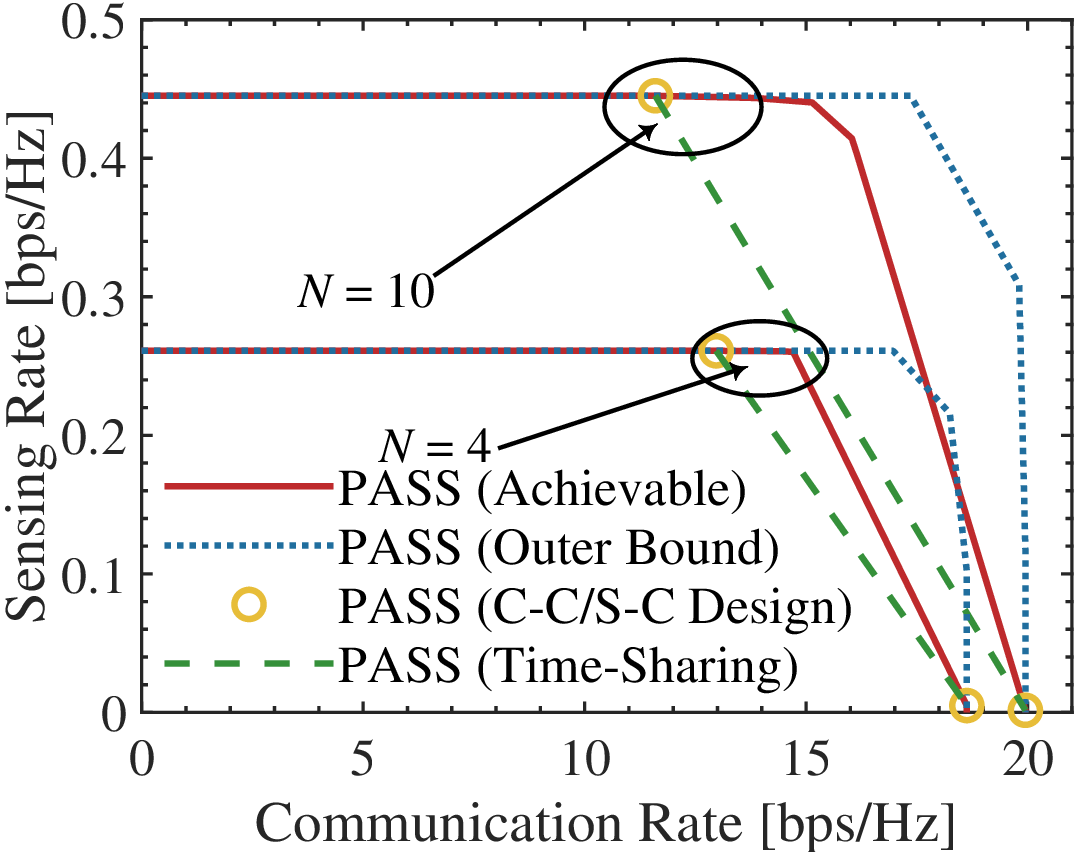}
	   \label{Figure_MP_Rate_Range_a}	
    }
   \subfigure[$x_{\rm{c}}=0$ m.]
    {
        \includegraphics[width=0.22\textwidth]{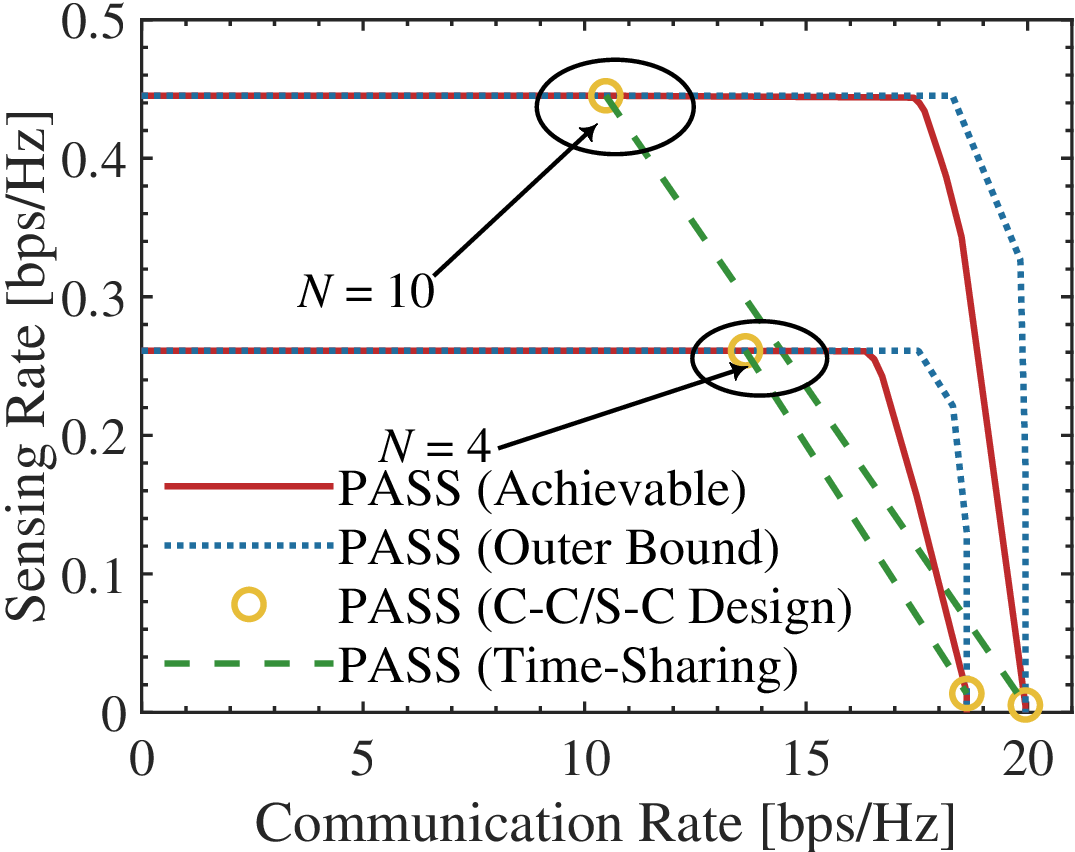}
	   \label{Figure_MP_Rate_Range_b}	
    }\\
    \subfigure[$x_{\rm{c}}=-3$ m.]
    {
        \includegraphics[width=0.22\textwidth]{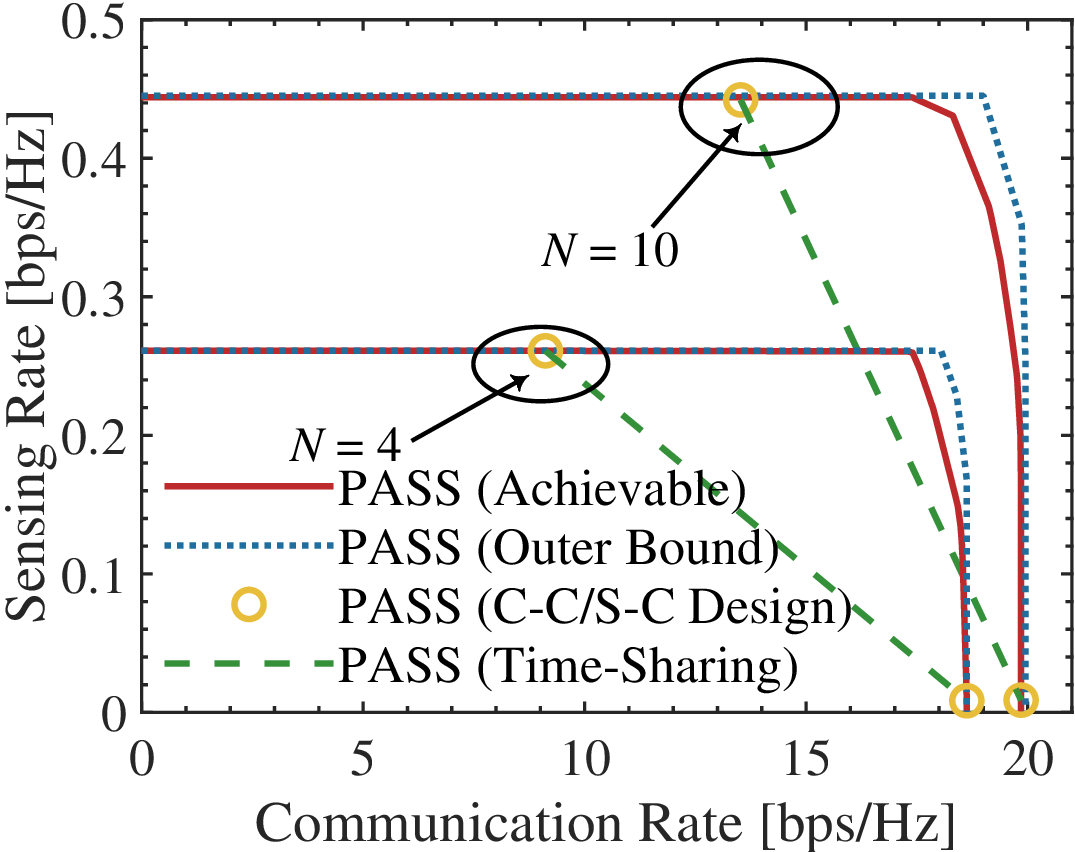}
	   \label{Figure_MP_Rate_Range_c}	
    }
   \subfigure[$x_{\rm{c}}=-8$ m.]
    {
        \includegraphics[width=0.22\textwidth]{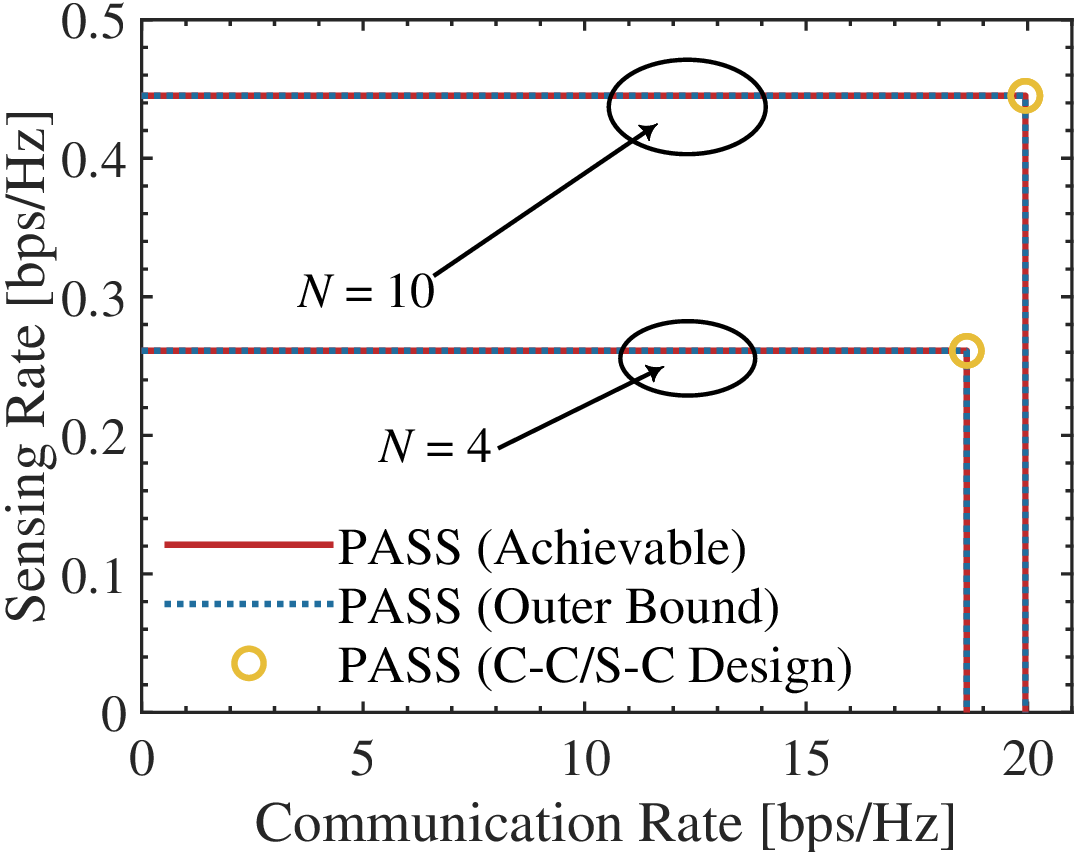}
	   \label{Figure_MP_Rate_Range_d}	
    }
\caption{Instantaneous rate region of multiple-pinch PASS (Case {\uppercase\expandafter{\romannumeral1}}). $y_{\rm{c}}=1$ m, $y_{\rm{s}}=-1$ m, and $x_{\rm{s}}=-8$ m.}
\label{Figure_MP_Rate_Range}
\vspace{-10pt}
\end{figure}

\subsubsection{Rate Region}
{\figurename} {\ref{Figure_MP_Rate_Range}} illustrates the \emph{instantaneous} rate region achieved by the multiple-pinch PASS under various values of $x_{\rm{c}}$, while fixing $y_{\rm{c}}=1$ m, $y_{\rm{s}}=-1$ m, and $x_{\rm{s}}=-8$ m. For comparison, we plot: \romannumeral1) the \emph{achievable inner bound} on the complete rate region (defined in \eqref{Outer_Bound_ISAC}), computed via the element-wise optimization method (Algorithm \ref{Algorithm1}) with time sharing, \romannumeral2) the \emph{outer bound} on the complete region (defined in \eqref{Outer_Bound_ISAC}), derived from the relaxation in \eqref{Upper_Bound_Cauchy–Schwarz}, \eqref{Extreme_Value_Lemma_Most_Important}, and \eqref{CS_Inequality_Result_Then}, and \romannumeral3) the rate region achieved through \emph{time sharing }between the C-C and S-C designs. To facilitate the discussion, results are presented for the ideal case (Case {\uppercase\expandafter{\romannumeral1}}) only. 

From {\figurename} {\ref{Figure_MP_Rate_Range}}, the inner and outer bounds closely align across all examined scenarios. Notably, as $x_{\rm{c}}$ approaches $x_{\rm{s}}$, the two bounds converge further, which indicates that the proposed bounds serve as accurate approximations of the complete rate region. Moreover, the rate region becomes increasingly rectangular as $x_{\rm{c}}\rightarrow x_{\rm{s}}$, which is consistent with the observations made in {\figurename} \ref{Figure_SP_Rate_Range}. We also observe that increasing the number of pinching antennas significantly enlarges the achievable rate region. Unlike the single-pinch case, we find that time sharing between C-C and S-C designs fails to approach the complete rate region in the multiple-pinch scenario. This underscores the necessity of optimized pinching beamforming to balance communication and sensing performance effectively.

\begin{figure}[!t]
\centering
\includegraphics[width=0.4\textwidth]{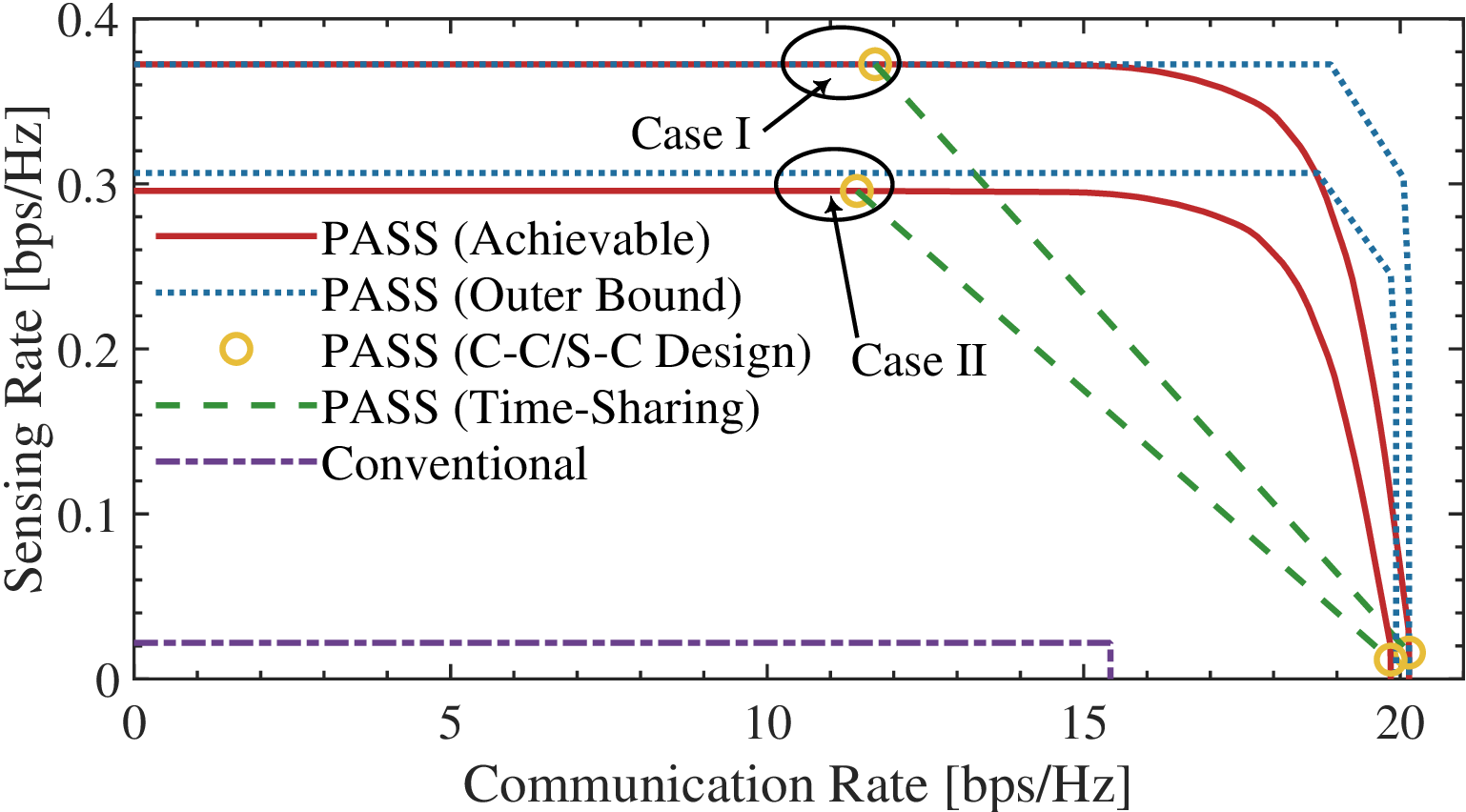}
\caption{Average rate region of multiple-pinch PASS. $N=10$.}
\label{Figure_MP_Average_Rate_Range}
\vspace{-10pt}
\end{figure}

{\figurename} \ref{Figure_MP_Average_Rate_Range} compares the \emph{average} rate region achieved by the multiple-pinch PASS against the conventional fixed-antenna system, by considering uniform distributions for both the communication user and sensing target. Results are presented for both ideal (Case {\uppercase\expandafter{\romannumeral1}}) and lossy (Case {\uppercase\expandafter{\romannumeral2}}) waveguide scenarios. For Case {\uppercase\expandafter{\romannumeral2}}, the outer bound is derived from \eqref{Upper_Bound_Cauchy–Schwarz} and \eqref{Extreme_Value_Lemma_Most_Important}, with in-waveguide propagation losses calculated based on the distance $\lvert t_0-\min\{x_{\rm{c}},x_{\rm{s}}\}\rvert$. The results demonstrate that time sharing between S-C and C-C designs fails to approximate the complete rate region for multiple-pinch PASS. This is consistent with the findings in {\figurename} {\ref{Figure_MP_Rate_Range}}. Furthermore, the multiple-pinch PASS achieves a substantially larger rate region than the conventional fixed-antenna system, regardless of whether in-waveguide propagation loss is considered. 

The above observations confirms the capability of PASS to enhance ISAC performance and suggests that increasing the number of pinching antennas is an effective means of expanding the achievable rate region.
\section{Conclusion}\label{Section_Conclusion}
This article analyzed the performance limits of PASS-assisted ISAC systems in terms of the achievable CR and SR. We characterized the impact of dual-functional pinching beamforming on the trade-off between communications and sensing by deriving the corresponding CR-SR rate region. For the single-pinch case, we obtained this rate region by solving a rate-profile optimization problem and applying a time-sharing strategy. For the multiple-pinch case, we derived inner and outer bounds on the rate region using an alternating optimization method along with several relaxation techniques. Theoretical analysis and numerical simulations confirmed that PASS achieves a larger CR-SR rate region and offers a more favorable CR-SR trade-off than conventional fixed-antenna systems, even when in-waveguide propagation loss is taken into account. We also observed that increasing the number of pinching antennas further amplifies this advantage. Moreover, our results showed that the time-sharing strategy effectively approximates the full rate region in the single-pinch scenario. In contrast, for the multiple-pinch case, achieving a desirable CR-SR trade-off or a promising rate region requires the careful design of the pinching beamforming strategy.
\begin{appendix}
\subsection{Proof of Lemma \ref{Lemma_SR}}\label{Proof_Lemma_SR}
Given $\{{\phi}_{\rm{r}},{{g}}_{\rm{r}},{\mathbf{h}}({\mathbf{u}}_{\rm{s}},{\mathbf{t}}),{\bm\phi}_{\rm{t}}\}$, the echo signal in \eqref{Sensing Model} can be rewritten as follows:
\begin{equation}\label{Sensing_Model_MSE}
\begin{split}
{\phi}_{\rm{r}}{\mathbf{y}}_{\rm{s}}^{\mathsf{T}}=\sqrt{{P}/{N}}{\phi}_{\rm{r}}{{g}}_{\rm{r}}{\bm\phi}_{\rm{t}}^{\mathsf{T}}{\mathbf{h}}({\mathbf{u}}_{\rm{s}},{\mathbf{t}})
{\mathbf{s}}^{\mathsf{T}}{\beta}_{\rm{RCS}}
+{\phi}_{\rm{r}}{\mathbf{n}}_{\rm{s}}^{\mathsf{T}}\triangleq {\mathbf{y}}_{{\rm{v}}}.
\end{split}
\end{equation}
It follows that the conditional MI between ${\mathbf{y}}_{{\rm{v}}}={\phi}_{\rm{r}}{\mathbf{y}}_{\rm{s}}^{\mathsf{T}}$ and ${\beta}_{\rm{RCS}}$ is equivalent to the capacity of the following virtual single-input multiple-output (SIMO) Gaussian channel with Gaussian-distributed input ${\beta}_{\rm{RCS}}\sim{\mathcal{CN}}(0,\alpha_{\rm{s}})$: 
\begin{align}
{\mathbf{y}}_{{\rm{v}}}={\mathbf{h}}_{\rm{v}}{\beta}_{\rm{RCS}} +{\mathbf{n}}_{\rm{v}},
\end{align}
where ${\mathbf{h}}_{\rm{v}}\triangleq\sqrt{{P}/{N}}{\phi}_{\rm{r}}{{g}}_{\rm{r}}{\bm\phi}_{\rm{t}}^{\mathsf{T}}{\mathbf{h}}({\mathbf{u}}_{\rm{s}},{\mathbf{t}})
{\mathbf{s}}^{\mathsf{T}}\in{\mathbbmss{C}}^{L\times1}$ denotes the virtual channel vector, and ${\mathbf{n}}_{\rm{v}}\triangleq{\phi}_{\rm{r}}{\mathbf{n}}_{\rm{s}}^{\mathsf{T}}\sim{\mathcal{CN}}({\mathbf{0}},\sigma_{\rm{s}}^2{\mathbf{I}}_L)$ represents the Gaussian noise. Accordingly, the sensing MI is given by \cite{el2011network}
\begin{subequations}
\begin{align}
I({\phi}_{\rm{r}}{\mathbf{y}}_{\rm{s}};{\mathbf{g}}|{\mathbf{x}})&=I({\mathbf{y}}_{{\rm{v}}};{\beta}_{\rm{RCS}}|{\mathbf{h}}_{\rm{v}})\\ 
&=\log _2\det ( \mathbf{I}_{L}+\alpha_{\rm{s}}{\mathbf{h}}_{\rm{v}}{\mathbf{h}}_{\rm{v}}^{\mathsf{H}}/\sigma_{\rm{s}}^2 ).
\end{align}
\end{subequations}
By applying Sylvester's identity, this expression simplifies to
\begin{subequations}\label{sensing_MI}
\begin{align}
I({\phi}_{\rm{r}}{\mathbf{y}}_{\rm{s}};{\mathbf{g}}|{\mathbf{x}})&=\log _2( 1+\alpha_{\rm{s}}/\sigma_{\rm{s}}^2{\mathbf{h}}_{\rm{v}}^{\mathsf{H}}{\mathbf{h}}_{\rm{v}} )\\
 &=\log _2\left( 1+\frac{P\alpha_{\rm{s}}\rvert{{g}}_{\rm{r}}\rvert^2
 \lvert{\mathbf{h}}^{\mathsf{T}}({\mathbf{u}}_{\rm{s}},{\mathbf{t}})
{\bm\phi}_{\rm{t}}\rvert^2\lVert{\mathbf{s}}\rVert^2}{\sigma_{\rm{s}}^2N/\lvert{\phi}_{\rm{r}}\rvert^2} \right).  
\end{align}
\end{subequations}
Given that $\lvert{\phi}_{\rm{r}}\rvert^2=1$ and $L^{-1}\lVert{\mathbf{s}}\rVert^2=1$, this result yields the expression shown in \eqref{Lemma_SR_Exp}.
\subsection{Proof of Lemma \ref{Lemma_MSE_SR}}\label{Proof_MSE_SR}
Referring to \eqref{Sensing_Model_MSE}, the MSE for an estimate $f_{\rm{est}}({\mathbf{y}}_{\rm{v}})$ of the RCS ${\beta}_{\rm{RCS}}$, based on the observation ${\mathbf{y}}_{\rm{v}}$, is given by
\begin{align}
{\mathsf{MSE}}={\mathbbmss{E}}\{\lvert f_{\rm{est}}({\mathbf{y}}_{\rm{v}})-{\beta}_{\rm{RCS}}\rvert^2\}.\label{MSE_Def}
\end{align}
It is well known that the minimum value of \eqref{MSE_Def} is achieved by the conditional mean estimator, given by \cite{kay1993fundamentals}
\begin{align}
f_{\rm{est}}({\mathbf{y}}_{\rm{v}})={\mathbbmss{E}}\{{\beta}_{\rm{RCS}}|{\mathbf{y}}_{\rm{v}}\}.
\end{align}
To compute this conditional mean, the conditional probability density function (PDF) $f_{{\beta}_{\rm{RCS}}|{\mathbf{y}}_{\rm{v}}}(x|{\mathbf{y}})$ is required. According to \emph{Bayes' theorem} \cite{kay1993fundamentals}, this is given by
\begin{align}\label{PDF_MSE_Bayes}
f_{{\beta}_{\rm{RCS}}|{\mathbf{y}}_{\rm{v}}}(x|{\mathbf{y}})=
\frac{f_{{\beta}_{\rm{RCS}}}(x)f_{{\mathbf{y}}_{\rm{v}}|{\beta}_{\rm{RCS}}}({\mathbf{y}}|x)}{f_{{\mathbf{y}}_{\rm{v}}}({\mathbf{y}})},
\end{align}
where $f_{{\beta}_{\rm{RCS}}}(\cdot)$, $f_{{\mathbf{y}}_{\rm{v}}}(\cdot)$, and $f_{{\mathbf{y}}_{\rm{v}}|{\beta}_{\rm{RCS}}}(\cdot|\cdot)$ denote the PDFs of ${\beta}_{\rm{RCS}}$, ${\mathbf{y}}_{\rm{v}}$, and ${\mathbf{y}}_{\rm{v}}$ conditioned on ${\beta}_{\rm{RCS}}$, respectively. Given ${\beta}_{\rm{RCS}}\sim{\mathcal{CN}}(0,\alpha_{\rm{s}})$ and ${\mathbf{n}}_{\rm{v}}\sim{\mathcal{CN}}({\mathbf{0}},\sigma_{\rm{s}}^2{\mathbf{I}}_L)$, we have
\begin{subequations}\label{PDF_Used_MSE}
\begin{align}
f_{{\beta}_{\rm{RCS}}}(x)&=\frac{1}{\pi \alpha_{\rm{s}}}{\rm{e}}^{-\frac{|x|^2}{\alpha_{\rm{s}}}},\\
f_{{\mathbf{y}}_{\rm{v}}|{\beta}_{\rm{RCS}}}({\mathbf{y}}|x)&=\frac{1}{(\pi \sigma_{\rm{s}}^2)^{L}}
{\rm{e}}^{-\frac{1}{\sigma_{\rm{s}}^2}({\mathbf{y}}-{\mathbf{h}}_{\rm{v}}x)^{\mathsf{H}}({\mathbf{y}}-{\mathbf{h}}_{\rm{v}}x)},\\
f_{{\mathbf{y}}_{{\rm{v}}}}({\mathbf{y}})&=
\frac{{\rm{e}}^{-{\mathbf{y}}^{\mathsf{H}}(\sigma_{\rm{s}}^2{\mathbf{I}_{L}}
+\alpha_{\rm{s}}{\mathbf{h}}_{\rm{v}}{\mathbf{h}}_{\rm{v}}^{\mathsf{H}})^{-1}{\mathbf{y}}}}{\pi^{L}\det(\sigma_{\rm{s}}^2{\mathbf{I}_{L}}
+\alpha_{\rm{s}}{\mathbf{h}}_{\rm{v}}{\mathbf{h}}_{\rm{v}}^{\mathsf{H}})}.
\end{align}
\end{subequations}
Substituting \eqref{PDF_Used_MSE} into \eqref{PDF_MSE_Bayes}, and using the identities $(\sigma_{\rm{s}}^2{\mathbf{I}_{L}}
+\alpha_{\rm{s}}{\mathbf{h}}_{\rm{v}}{\mathbf{h}}_{\rm{v}}^{\mathsf{H}})^{-1}=\frac{1}{\sigma_{\rm{s}}^2}({\mathbf{I}_{L}}-
\frac{\alpha_{\rm{s}}}{\sigma_{\rm{s}}^2+\alpha_{\rm{s}}\lVert{\mathbf{h}}_{\rm{v}}\rVert^2}
{\mathbf{h}}_{\rm{v}}{\mathbf{h}}_{\rm{v}}^{\mathsf{H}})$ and $\det(\sigma_{\rm{s}}^2{\mathbf{I}_{L}}
+\alpha_{\rm{s}}{\mathbf{h}}_{\rm{v}}{\mathbf{h}}_{\rm{v}}^{\mathsf{H}})=(\sigma_{\rm{s}}^2)^{L}(1+\frac{\alpha_{\rm{s}}}{\sigma_{\rm{s}}^2}\lVert{\mathbf{h}}_{\rm{v}}\rVert^2)$, we obtain the following \emph{posterior} distribution of ${\beta}_{\rm{RCS}}$ given ${\mathbf{y}}_{\rm{v}}$:
\begin{align}
f_{{\beta}_{\rm{RCS}}|{\mathbf{y}}_{\rm{v}}}(x|{\mathbf{y}})=
\frac{1}{\pi \alpha_{\rm{s}}\frac{1}{\omega_{\rm{s}}}\sigma_{\rm{s}}^2}{\rm{e}}^{-\frac{\omega_{\rm{s}}}{\alpha_{\rm{s}}\sigma_{\rm{s}}^2}\lvert x - \frac{\alpha_{\rm{s}}}{\omega_{\rm{s}}}{{\mathbf{h}}_{\rm{v}}^{\mathsf{H}}{\mathbf{y}}}\rvert^2},
\end{align}
where $\omega_{\rm{s}}\triangleq{\sigma_{\rm{s}}^2}+{\alpha_{\rm{s}}}\lVert{\mathbf{h}}_{\mathsf{v}}\rVert^2$. Thus, the conditional mean estimator becomes
\begin{align}
{\mathbbmss{E}}\{{\beta}_{\rm{RCS}}|{\mathbf{y}}_{\rm{v}}\}=\int_{\mathbbmss{C}}xf_{{\beta}_{\rm{RCS}}|{\mathbf{y}}_{\rm{v}}}(x|{\mathbf{y}}){\rm{d}}x
=\frac{\alpha_{\rm{s}}}{\omega_{\rm{s}}}{{\mathbf{h}}_{\rm{v}}^{\mathsf{H}}{\mathbf{y}}_{\rm{v}}}.
\end{align}
This result implies that the conditional mean estimate of the RCS ${\beta}_{\rm{RCS}}$ is a linear function of the observation ${\mathbf{y}}_{\rm{v}}$, i.e.,
\begin{align}\label{MMSE_Estimator}
f_{\rm{est}}({\mathbf{y}}_{\rm{v}})={\mathbbmss{E}}\{{\beta}_{\rm{RCS}}|{\mathbf{y}}_{\rm{v}}\}=
\frac{\alpha_{\rm{s}}}{{\sigma_{\rm{s}}^2}+{\alpha_{\rm{s}}}\lVert{\mathbf{h}}_{\rm{v}}\rVert^2}{{\mathbf{h}}_{\rm{v}}^{\mathsf{H}}{\mathbf{y}}_{\rm{v}}}.
\end{align}
Substituting \eqref{Sensing_Model_MSE} and \eqref{MMSE_Estimator} into \eqref{MSE_Def} gives
\begin{subequations}\label{MSE_Expression_Step2}
\begin{align}
{\mathsf{MSE}}&={\mathbbmss{E}}\left\{
\left\lvert
\left(\frac{\alpha_{\rm{s}}\lVert{\mathbf{h}}_{\rm{v}}\rVert^2}{\omega_{\rm{s}}}-1\right){\beta}_{\rm{RCS}}
+\frac{\alpha_{\rm{s}}{{\mathbf{h}}_{\rm{v}}^{\mathsf{H}}{\mathbf{n}}_{\rm{v}}}}{\omega_{\rm{s}}}
\right\rvert^2
\right\}\\
&={\mathbbmss{E}}\left\{
\left\lvert
\frac{-{\sigma_{\rm{s}}^2}{\beta}_{\rm{RCS}}}{{\sigma_{\rm{s}}^2}+{\alpha_{\rm{s}}}\lVert{\mathbf{h}}_{\rm{v}}\rVert^2}
+\frac{\alpha_{\rm{s}}{{\mathbf{h}}_{\rm{v}}^{\mathsf{H}}{\mathbf{n}}_{\rm{v}}}}{{\sigma_{\rm{s}}^2}+{\alpha_{\rm{s}}}\lVert{\mathbf{h}}_{\rm{v}}\rVert^2}
\right\rvert^2
\right\}.
\end{align}
\end{subequations}
Since ${\beta}_{\rm{RCS}}$ and ${\mathbf{n}}_{\rm{v}}$ are independent, the MSE simplifies to
\begin{align}
{\mathsf{MSE}}=\frac{\alpha_{\mathrm{s}}\sigma_{\rm{s}}^2}{\sigma_{\rm{s}}^2+\alpha_{\rm{s}}\lVert{\mathbf{h}}_{\mathsf{v}}\rVert^2}
=\frac{\alpha_{\mathrm{s}}}{\frac{P\alpha_{\rm{s}}\rvert{{g}}_{\rm{r}}\rvert^2
 \lvert{\mathbf{h}}^{\mathsf{T}}({\mathbf{u}}_{\rm{s}},{\mathbf{t}})
{\bm\phi}_{\rm{t}}\rvert^2\lVert{\mathbf{s}}\rVert^2}{\sigma_{\rm{s}}^2N/\lvert{\phi}_{\rm{r}}\rvert^2}}.\label{MSE_Formulation}
\end{align}
Comparing this result with the sensing MI in \eqref{sensing_MI} confirms that the pinching beamformer $\mathbf{t}$ that maximizes the SR also minimizes the MSE in estimating the RCS ${\beta}_{\rm{RCS}}$. In other words, maximizing the sensing rate is equivalent to minimizing the estimation error of ${\beta}_{\rm{RCS}}$, which concludes the proof.
\subsection{Proof of Theorem \ref{Theorem_SP_Pareto}}\label{Proof_Theorem_SP_Pareto}
We begin by showing that the Pareto-optimal activated location must lie within the interval $[x_{\rm{c}},x_{\rm{s}}]$. This can be demonstrated via contradiction. If $t_1<x_{\rm{c}}$, then we can update $t_1\leftarrow2x_{\rm{c}}-t_{1}$, which maintains the same CR but yields a higher SR. Similarly, if $t_1>x_{\rm{s}}$, then updating $t_1\leftarrow2x_{\rm{s}}-t_{1}$ preserves the SR while improving the CR. In both cases, the rate pair $({\mathcal{R}}_{\rm{c}},{\mathcal{R}}_{\rm{s}})$ can be improved, implying that such locations cannot correspond to Pareto-optimal solutions. Therefore, the optimal activated location must satisfy $t_1\in[x_{\rm{c}},x_{\rm{s}}]$. 

We then parameterize the activated location as $t_1=x_{\rm{c}}+\beta\Delta_x$ with $\beta\in[0,1]$ and $\Delta_x=\lvert x_{\rm{c}}-x_{\rm{s}}\rvert$. From \eqref{SP_Pareto}, the rate-profile problem seeks to maximize the following:
\begin{align}
{\mathcal{R}}_{\alpha}=\max_{\beta\in[0,1]}\min\left\{{\mathcal{R}}_{\rm{c}}/{\alpha},{\mathcal{R}}_{\rm{s}}/({1-\alpha})\right\}.
\end{align}
For convenience, we define 
\begin{align}
&\frac{{\mathcal{R}}_{\rm{c}}}{\alpha}=\frac{1}{\alpha}\log_2\left(1+\frac{\overline{\gamma}_{\rm{c}}}
{d_{\rm{c}}^2+\beta^2\Delta_x^2}\right)\triangleq f_{\rm{c}}(\beta),\\
&\frac{{\mathcal{R}}_{\rm{s}}}{1-\alpha}=\frac{1}{L(1-\alpha)}\log_2\left(1+\frac{\overline{\gamma}_{\rm{s}}}
{d_{\rm{s}}^2+(1-\beta)^2\Delta_x^2}\right)\triangleq f_{\rm{s}}(\beta).
\end{align}
It is easy to verify that $f_{\rm{c}}(\beta)$ (or $f_{\rm{s}}(\beta)$) is strictly decreasing (or increasing) with $\beta\in[0,1]$. Let $\beta_{\alpha}^{\star}$ denote the optimal $\beta$. Then, if $f_{\rm{c}}(1)\geq f_{\rm{s}}(1)$, we have ${\mathcal{R}}_{\alpha}=f_{\rm{s}}(1)$ and $\beta_{\alpha}^{\star}=1$. If $f_{\rm{c}}(0)\leq f_{\rm{s}}(0)$, we have ${\mathcal{R}}_{\alpha}=f_{\rm{c}}(0)$ and $\beta_{\alpha}^{\star}=0$. Otherwise, ${\mathcal{R}}_{\alpha}$ is determined by the point where $f_{\rm{c}}(\beta)=f_{\rm{s}}(\beta)$ for $\beta\in[0,1]$, i.e., the intersection of the two functions. In this case, ${\mathcal{R}}_{\alpha}=f_{\rm{c}}(\beta_{\alpha}^{\star})=f_{\rm{s}}(\beta_{\alpha}^{\star})$, and the optimal $\beta_{\alpha}^{\star}$ can be obtained using a bisection search over the interval $\beta\in[0,1]$ to solve the equation $f_{\rm{c}}(\beta)=f_{\rm{s}}(\beta)$. This concludes the proof.
\subsection{Proof of Theorem \ref{Theorem_SP_Capacity_Region}}\label{Proof_Theorem_SP_Capacity_Region}
When $x_{\rm{f}}<x_{\rm{c}}$, setting $t_1=x_{\rm{c}}$ results in ${\mathcal{R}}_{\rm{c}}>{\mathcal{R}}_{\rm{c}}^{\rm{f}}$ and ${\mathcal{R}}_{\rm{s}}>{\mathcal{R}}_{\rm{s}}^{\rm{f}}$. Similarly, if $x_{\rm{f}}>x_{\rm{s}}$, setting $t_1=x_{\rm{s}}$ leads to ${\mathcal{R}}_{\rm{c}}>{\mathcal{R}}_{\rm{c}}^{\rm{f}}$ and ${\mathcal{R}}_{\rm{s}}>{\mathcal{R}}_{\rm{s}}^{\rm{f}}$. In the case where $x_{\rm{c}}<x_{\rm{f}}<x_{\rm{s}}$, setting $t_1=x_{\rm{f}}$ leads to $({\mathcal{R}}_{\rm{c}},{\mathcal{R}}_{\rm{s}})=({\mathcal{R}}_{\rm{c}}^{\rm{f}},{\mathcal{R}}_{\rm{s}}^{\rm{f}})$. These observations confirm that, for any $x_{\rm{f}}$, the rate pair $({\mathcal{R}}_{\rm{c}}^{\rm{f}},{\mathcal{R}}_{\rm{s}}^{\rm{f}})$ always falls within the region ${\mathcal{C}}_{\rm{p}}^{\rm{S}}$. Therefore, we conclude that ${\mathcal{C}}_{\rm{f}}\subseteq{\mathcal{C}}_{\rm{p}}^{\rm{S}}$.
\subsection{Proof of Lemma \ref{Extreme_Value_Lemma}}\label{Proof_Extreme_Value_Lemma}
The proof of Lemma \ref{Extreme_Value_Lemma} builds on \emph{Bauer's maximum principle} \cite{bauer1958minimalstellen}, which states that any convex and continuous function defined over a convex and compact set attains its maximum at an extreme point of that set. Specifically, since the function $\frac{1}{{\rho^2+\beta_n^2}}$ is strictly convex over $\beta_n^2\in[0,1]$ for $n\in{\mathcal{N}}$, Bauer's principle ensures that the sum $\sum_{n=1}^{N}\frac{1}{{\rho^2+\beta_n^2}}$ is maximized when each $\beta_n^2$ is set to either $0$ or $1$, with at most one variable taking an intermediate value if $S_{\bm\beta}=\sum_{n=1}^{N}\beta_n^2$ (i.e., the total sum) is not an integer. A more rigorous proof relies on \emph{Karamata's inequality} \cite{kadelburg2005inequalities}, which is stated as follows.
\subsubsection*{Karamata's inequality}
Let $\mathcal{I}\subseteq{\mathbbmss{R}}$ be an interval of the real line, and let $f_{\rm{cx}}(\cdot)$ be a real-valued convex function defined on $\mathcal{I}$. Suppose $\mathring{x}_1,\ldots,\mathring{x}_k$ and $\mathring{y}_1,\ldots,\mathring{y}_k$ are elements of $\mathcal{I}$ such that $\mathring{x}_1,\ldots,\mathring{x}_k$ \emph{majorizes} $\mathring{y}_1,\ldots,\mathring{y}_k$. Then,
\begin{align}\label{K_Inequality}
f_{\rm{cx}}(\mathring{x}_1)+\ldots+f_{\rm{cx}}(\mathring{x}_k)\geq f_{\rm{cx}}(\mathring{y}_1)+\ldots+f_{\rm{cx}}(\mathring{y}_k),
\end{align}
where majorization means the following:
\begin{subequations}
\begin{align}
&\mathring{x}_1\geq \mathring{x}_2\geq\ldots\geq \mathring{x}_k,\\
&\mathring{y}_1\geq \mathring{y}_2\geq\ldots\geq \mathring{y}_k,\\
&\mathring{x}_1+\ldots+\mathring{x}_i\geq \mathring{y}_1+\ldots+\mathring{y}_i,~i\in\{1,\ldots,k-1\},\\
&\mathring{x}_1+\ldots+\mathring{x}_k= \mathring{y}_1+\ldots+\mathring{y}_k.
\end{align}
\end{subequations}
If $f_{\rm{cx}}(\cdot)$ is strictly convex, the inequality in \eqref{K_Inequality} becomes strict unless $\mathring{x}_i=\mathring{y}_i$ for all $i\in\{1,\ldots,k\}$.

We then apply this inequality to our case. Given $\beta_1^2+\ldots+\beta_N^2=S_{\bm\beta}$, it follows that the sequence
\begin{align}
\underbrace{1,\ldots,1}_{\lfloor{S}_{\bm\beta}\rfloor~{\text{ones}}},S_{\bm\beta}-\lfloor{S}_{\bm\beta}\rfloor,
\underbrace{0,\ldots,0}_{N-\lfloor{S}_{\bm\beta}\rfloor-{\mathbbmss{1}}_{S_{\bm\beta}\in{\mathbbmss{Z}}}~{\text{zeros}}}
\end{align}
\emph{majorizes} any $\beta_1^2,\ldots,\beta_N^2$ satisfying $\beta_n^2\in[0,1]$ ($n\in{\mathcal{N}}$) and $\sum_{n=1}^{N}\beta_n^2=S_{\bm\beta}$. Combining this with \eqref{K_Inequality} leads to \eqref{Extreme_Value_Lemma_Most_Important}.
\end{appendix}
\bibliographystyle{IEEEtran}
\bibliography{mybib}
\end{document}